\documentclass[prd,aps,twocolumn,preprintnumbers,amsmath,amssymb,floatfix,raggedbottom]{revtex4-2}
\usepackage{graphicx}
\usepackage[breaklinks,colorlinks=true,linkcolor=blue,citecolor=blue,urlcolor=blue]{hyperref}
\usepackage{bm}
\usepackage{physics}

\newcommand{\Mcl}{M_{\mathrm{cl}}}
\newcommand{\occ}{\mathrm{occ}}
\newcommand{\kp}{\bm k_\perp}
\newcommand{\Dp}{\bm\Delta_\perp}
\newcommand{\pket}{\bm p_{\mathrm{ket}}}
\newcommand{\cA}{\mathcal A}
\newcommand{\cV}{\mathcal V}
\newcommand{\tcS}{\widetilde{\mathcal S}}
\newcommand{\tcV}{\widetilde{\mathcal V}}
\newcommand{\Dop}{\hat{\mathcal D}}
\newcommand{\cS}{\mathcal S}
\newcommand{\upd}{{u+d}}
\newcommand{\umd}{{u-d}}
\newcommand{\upmd}{{u\pm d}}
\graphicspath{}

\begin{document}

\title{Proton GTMDs and Wigner Distributions in the Chiral Quark Soliton Model: \\ Orbital Angular Momentum and Spin--Orbit Correlation with Sea Quarks}

\author{Josuke Minamiguchi}
\email{minamiguchi@nt.phys.s.u-tokyo.ac.jp}
\affiliation{Department of Physics, The University of Tokyo, 7-3-1 Hongo, Bunkyo-ku, Tokyo 113-0033, Japan}

\author{Tomoya Uji}
\email{uji@nt.phys.s.u-tokyo.ac.jp}
\affiliation{Department of Physics, The University of Tokyo, 7-3-1 Hongo, Bunkyo-ku, Tokyo 113-0033, Japan}

\begin{abstract}
We calculate the leading-twist vector and axial-vector quark generalized transverse-momentum-dependent distributions (GTMDs) of the proton in the chiral quark soliton model, including both the discrete valence level and the regularized Dirac sea.
We consider the isosinglet and isotriplet combinations of GTMDs at zero and nonzero skewness.
At zero skewness, we construct Wigner distributions and resolve the orbital angular momentum (OAM) and spin--orbit correlation (SOC) densities in transverse position and momentum space, separately for quarks and antiquarks.  This provides a phase-space picture of the sea-quark and sea-antiquark contributions, displaying OAM aligned with the proton spin and anti-aligned with the partons' own spin.
\end{abstract}

\maketitle

\section{Introduction}\label{sec:intro}
Generalized transverse-momentum-dependent distributions (GTMDs) provide a unified framework for describing the multidimensional partonic structure of hadrons~\cite{Meissner:2009ww}.  Specific moments of GTMDs give generalized parton distributions (GPDs)~\cite{Muller:1994ses, Ji:1996ek, Radyushkin:1996nd, Diehl:2003ny} and transverse-momentum-dependent parton distributions (TMDs)~\cite{Mulders:1995dh, Bacchetta:2006tn, Boussarie:2023izj}, which encode complementary information on the transverse spatial and momentum distributions of partons.  GTMDs also retain correlations between transverse position and momentum that are lost in the separate GPD and TMD projections.  At zero skewness, a Fourier transform of the GTMDs with respect to the transverse momentum transfer yields Wigner distributions, which depend on the longitudinal momentum fraction, $x$, transverse impact parameter, $\bm{b}_\perp$, and transverse momentum, $\kp$~\cite{Ji:2003ak, Belitsky:2003nz, Lorce:2011kd}.  These phase-space distributions provide a framework for investigating parton orbital angular momentum (OAM) and spin--orbit correlation (SOC)~\cite{Lorce:2011kd,  Hatta:2011ku, Lorce:2014mxa}.

Progress toward accessing GTMDs has been made through several approaches.  Proposed experimental probes include diffractive dijet production for small-$x$ gluon GTMDs~\cite{Hatta:2016dxp} and exclusive double Drell--Yan scattering for quark GTMDs~\cite{Bhattacharya:2017bvs}.  More recently, exclusive meson electroproduction has been proposed as a probe of $\kp$-moments of the quark GTMDs~\cite{Bhattacharya:2023hbq} and the gluon GTMDs~\cite{Bhattacharya:2026qnd}.  Complementary lattice QCD calculations have evaluated quark OAM using GTMD matrix elements~\cite{Engelhardt:2017miy, Engelhardt:2020qtg}.  The perturbative description has also advanced through one-loop matching of leading-twist GTMDs onto GPDs~\cite{Bertone:2025vgy}.  Moreover, recent theoretical work has considered the small-$x$ regime for GTMDs~\cite{Benic:2026idy}, OAM~\cite{Hatta:2016aoc, Boussarie:2019icw, Kovchegov:2023yzd, Manley:2024pcl, Kovchegov:2026lik}, and SOC~\cite{Bhattacharya:2024sno, Benic:2025xpg}.  These developments motivate nonperturbative calculations of GTMDs.

Quark GTMDs and Wigner distributions have been investigated in several hadronic models, including light-front quark models~\cite{Lorce:2011dv, Luan:2024nwc} and the bag model~\cite{Maynard:2026atm}.  Earlier calculations in the chiral quark soliton model (CQSM) employed a light-front description restricted to the three-quark Fock sector~\cite{Lorce:2011dv}.  Chiral dynamics can generate a nonperturbative sea with a transverse-momentum structure substantially different from that of valence quarks~\cite{Schweitzer:2012hh}.  Extending the CQSM calculation to include the full contribution from the Dirac continuum therefore provides a way to determine how the nonperturbative sea modifies quark phase-space correlations and contributes to OAM and SOC.

The CQSM provides a field-theoretical description of the proton in which constituent quarks interact with a chiral field~\cite{Diakonov:1987ty, Reinhardt:1988fz, Alkofer:1994ph, Christov:1995vm}.
In this model, parton distributions receive contributions from both a discrete valence level and the negative-energy Dirac continuum distorted by this field.
The continuum restores the positivity of the isosinglet unpolarized antiquark distribution, which is violated if only the discrete level is retained~\cite{Diakonov:1997vc}.  The CQSM has been extensively used to study parton distribution functions (PDFs)~\cite{Weigel:1996kw, Diakonov:1996sr, Weigel:1996jh, Diakonov:1997vc, Weiss:1997rt, Wakamatsu:1997en, Pobylitsa:1998tk, Wakamatsu:1998rx, Weigel:1999pc, Wakamatsu:1999nj, Takyi:2019ahv}, GPDs~\cite{Petrov:1998kf, Penttinen:1999th, Ossmann:2004bp, Wakamatsu:2005vk}, and TMDs~\cite{Pobylitsa:2003ty, Wakamatsu:2009fn, Schweitzer:2012hh}.  Recent developments include calculations of twist-3 GPDs~\cite{Kim:2023pll}, chiral-odd GPDs~\cite{Kim:2024ibz, Kim:2025mol}, polarized PDFs with flavor-singlet mesons~\cite{Minamiguchi:2026uaq}, and unpolarized GPDs at nonzero skewness and momentum transfer with explicit sums over the Dirac sea~\cite{Minamiguchi:2026qcb}.

In this work, we calculate the leading-twist quark GTMDs, $F_{1, n}$ and $G_{1, n}$ ($n = 1, \ldots, 4$), in the CQSM, including both the discrete valence level and the regularized Dirac sea.
We treat both the isosinglet and isotriplet combinations, at zero and at nonzero skewness.
At zero skewness, we study Wigner distributions, OAM, and SOC.
We find that, at the momentum fractions studied, the isosinglet OAM of both quarks and antiquarks points along the proton spin and opposite to their own spin.
The Dirac sea carries about $60\%$ of the integrated isosinglet OAM, whereas the sea quarks and sea antiquarks give nearly opposite contributions to the SOC.

\section{Quark GTMDs at large $N_c$}\label{sec:gtmd}
We first define the quark GTMDs and introduce our notation.  We then summarize their large-$N_c$ scaling in the isosinglet and isotriplet channels.

\subsection{Definition of quark GTMDs}\label{sec:definitions}

We write the light-cone components of a four-vector, $v$, as $v^\pm = (v^0 \pm v^3) / \sqrt{2}$.  We write the transverse components of $v$ as $\bm v_\perp = (v^1, v^2)$.  The quark GTMDs of the proton are defined by the correlator~\cite{Meissner:2009ww}
\begin{equation}
\begin{split}
    &W^{[\Gamma],q}_{S'S} = \frac{1}{2}\int\frac{dz^-d^2\bm{z}_\perp}{(2\pi)^3}\,e^{ik\cdot z} \\
    &\qquad\times\mel**{P', S'}{\bar{q}\left(-\frac{z}{2}\right)\Gamma\mathcal{W}q\left(\frac{z}{2}\right)}{P, S}\eval_{z^+ = 0}\,.
    \label{eq:correlator}
\end{split}
\end{equation}
Here, $z$ is the separation between the two quark fields, $q$, and $\Gamma$ is a Dirac matrix.  The initial proton has the momentum, $P$, and the spin, $S$.  The final proton has the momentum, $P'$, and the spin, $S'$.  We write $\bar{P} = (P + P') / 2$ for the average proton momentum.  We write $\Delta = P' - P$ for the momentum transfer and define $t = \Delta^2$.  The average quark momentum, $k$, satisfies $k^+ = x\bar{P}^+$, and $\kp$ is the transverse part of $k$.  The skewness $\xi$ is defined as $\xi = - \Delta^+ / (2\bar{P}^+)$.  The Wilson line, $\mathcal{W}$, connects the two quark fields and ensures gauge invariance of the bilocal quark operator.

The correlator with $\Gamma = \gamma^+$ defines the unpolarized twist-2 GTMDs, $F_{1, n}$, as~\cite{Meissner:2009ww}
\begin{equation}
\begin{split}
    &W^{[\gamma^+], q}_{S'S} = \frac{1}{2M_N}\bar{u}(P', S')\biggl[F_{1, 1}^q + \frac{i\sigma^{i+}k_\perp^i}{\bar{P}^+}F_{1, 2}^q \\
    &\qquad + \frac{i\sigma^{i+}\Delta_\perp^i}{\bar{P}^+}F_{1, 3}^q + \frac{i\sigma^{ij}k_\perp^i\Delta_\perp^j}{M_N^2}F_{1, 4}^q\biggr]u(P, S)\,.
\label{eq:defF}
\end{split}
\end{equation}
The correlator with $\Gamma = \gamma^+\gamma_5$ defines the longitudinally polarized twist-2 GTMDs, $G_{1, n}$, as~\cite{Meissner:2009ww}:
\begin{equation}
\begin{split}
    &W^{[\gamma^+\gamma_5], q}_{S'S} = \frac{1}{2M_N}\bar{u}(P', S')\biggl[- \frac{i\epsilon^{-+ij}k_\perp^i\Delta_\perp^j}{M_N^2}G_{1, 1}^q \\
    &\qquad + \frac{i\sigma^{i+}\gamma_5k_\perp^i}{\bar{P}^+}G_{1, 2}^q + \frac{i\sigma^{i+}\gamma_5\Delta_\perp^i}{\bar{P}^+}G_{1, 3}^q \\
    &\qquad + i\sigma^{+-}\gamma_5G_{1, 4}^q\biggr]u(P, S)\,.
\label{eq:defG}
\end{split}
\end{equation}
Here, $M_N$ is the proton mass, $u(P, S)$ is the proton spinor, $\gamma_5 = i\gamma^0\gamma^1\gamma^2\gamma^3$, and $\sigma^{\alpha\beta} = \frac{i}{2}[\gamma^\alpha, \gamma^\beta]$.  The transverse indices, $i, j = 1, 2$, are summed over.  The Levi-Civita symbol is normalized as $\epsilon^{0123} = +1$, so that $\epsilon^{-+12} = +1$.  The GTMDs are functions of $x$, $\xi$, $|\kp|^2$, $\kp\cdot\Dp$, and $|\Dp|^2$~\cite{Meissner:2009ww}.  We suppress these arguments for brevity.

\subsection{Large-$N_c$ counting of GTMDs}\label{sec:spin}

We analyze the large-$N_c$ scaling of the quark GTMDs using the spin-flavor structure of the proton, following the corresponding analyses of GPDs and TMDs~\cite{Petrov:1998kf, Ossmann:2004bp, Pobylitsa:2003ty}.  We consider both the vector and axial-vector sectors and identify the leading and subleading isospin combinations.  The scaling obtained for the vector GTMDs agrees with the analysis in the bag model of Ref.~\cite{Maynard:2026atm}.

We evaluate the spinor bilinears in the Breit frame, where $\bar{\bm P}=0$ and $\Delta^0 = 0$.  We count $M_N = O(N_c)$ and $k, \bm{\Delta} = O(N_c^0)$ in this frame.  Both protons then have velocities of order $N_c^{-1}$, which makes the Breit frame convenient for the $1/N_c$ expansion of the spinor bilinears.  The kinematic variables satisfy $t = - |\bm{\Delta}|^2$ and $\xi = - \Delta^3/(2\bar{P}^0)$, so that $x, \xi = O(N_c^{-1})$ and $\bar{P}^0 = M_N[1 + O(N_c^{-2})]$.

Using proton spin states with spin quantized along the third axis, we decompose GTMD correlators into spin-independent and spin-dependent parts:
\begin{equation}
\begin{split}
    W^{[\gamma^+]\upmd}_{S'S} &= \delta_{S'S}\cS^\upmd + (\sigma^a)_{S'S}\cV^{\upmd, a}\,, \\
    W^{[\gamma^+\gamma_5]\upmd}_{S'S} &= \delta_{S'S}\tcS^\upmd + (\sigma^a)_{S'S}\tcV^{\upmd, a}\,,
\end{split}
\label{eq:Wdecomp}
\end{equation}
where $(\sigma^a)_{S'S} = \chi_{S'}^\dagger\sigma^a\chi_{S}$ and $\chi_{S}$ are two-component spinors.  The spin-flavor symmetry at large $N_c$ gives~\cite{Ossmann:2004bp, Pobylitsa:2003ty}
\begin{equation}
\begin{split}
    \cS^{\upd}, \tcS^{\upd}, \cV^{\umd, a}, \tcV^{\umd, a} &\sim O(N_c^2)\,, \\
    \cS^{\umd}, \tcS^{\umd}, \cV^{\upd, a}, \tcV^{\upd, a} &\sim O(N_c)\,.
\end{split}
\label{eq:spin_flavor_counting}
\end{equation}
Thus the spin-independent coefficients are leading in the isosinglet channel, while the spin-dependent coefficients are leading in the isotriplet channel.

We determine the coefficients of Eq.~\eqref{eq:Wdecomp} by evaluating the spinor bilinears and expanding their kinematic factors in $1/N_c$.
They then read
\begin{subequations}
\label{eq:coefficients}
\begin{align}
    \cS^\upd &= F^\upd_{1, 1}\,,
    \label{eq:Scoefupd} \\
    \cS^\umd &= F^\umd_{1, 1} - \frac{(\kp\cdot\Dp)F^\umd_{1, 2} + |\Dp|^2F^\umd_{1, 3}}{2M_N^2}\,,
    \label{eq:Scoefumd} \\
    \cV^{\upmd, i} &= \frac{i\epsilon^{-+ij}}{M_N}\left(k_\perp^jF^\upmd_{1, 2} + \Delta_\perp^jF^\upmd_{1, 3}\right)\,,
    \label{eq:Vcoef} \\
    \cV^{\upmd, 3} &= \frac{i\epsilon^{-+ij}}{M_N^2}k_\perp^i\Delta_\perp^jF^\upmd_{1,4}\,,
    \label{eq:V3coef} \\
    \tcS^\upd &= - \frac{i\epsilon^{-+ij}}{M_N^2}k_\perp^i\Delta_\perp^jG^\upd_{1, 1}\,,
    \label{eq:tScoefupd} \\
    \tcS^\umd &= - \frac{i\epsilon^{-+ij}}{M_N^2}k_\perp^i\Delta_\perp^j\left(G^\umd_{1, 1} - \frac{1}{2}G^\umd_{1, 2}\right)\,,
    \label{eq:tScoefumd} \\
    \tcV^{\upmd, i} &= \frac{1}{M_N}\left(k_\perp^iG^\upmd_{1, 2} + \Delta_\perp^iG^\upmd_{1, 3}\right)\,,
    \label{eq:tVcoef} \\
    \tcV^{\upmd, 3} &= G^\upmd_{1, 4}\,.
    \label{eq:tV3coef}
\end{align}
\end{subequations}
In Eqs.~\eqref{eq:Vcoef} and \eqref{eq:tVcoef}, $i = 1, 2$ denotes a transverse component.  Combining Eq.~\eqref{eq:coefficients} with the spin-flavor hierarchy in Eq.~\eqref{eq:spin_flavor_counting}, we obtain the GTMD scaling summarized in Table~\ref{tab:Nc}.  For each GTMD, we refer to the isospin combination with the larger power of $N_c$ as the leading combination and to the other as the subleading combination.

\begin{table}
\caption{Large-$N_c$ scaling of the quark GTMDs at fixed $N_cx$, $N_c\xi$, $\kp$, and $\Dp$.  The leading isospin combination for each GTMD is shown in bold.}
\label{tab:Nc}
\begin{ruledtabular}
\begin{tabular}{ccc}
    GTMD & $u+d$ & $u-d$ \\
    \hline
    $F_{1,1}$ & $\bm{N_c^2}$ & $N_c$ \\
    $F_{1,2}$ & $N_c^2$ & $\bm{N_c^3}$ \\
    $F_{1,3}$ & $N_c^2$ & $\bm{N_c^3}$ \\
    $F_{1,4}$ & $N_c^3$ & $\bm{N_c^4}$ \\
    $G_{1,1}$ & $\bm{N_c^4}$ & $N_c^3$ \\
    $G_{1,2}$ & $N_c^2$ & $\bm{N_c^3}$ \\
    $G_{1,3}$ & $N_c^2$ & $\bm{N_c^3}$ \\
    $G_{1,4}$ & $N_c$ & $\bm{N_c^2}$ \\
\end{tabular}
\end{ruledtabular}
\end{table}

\section{GTMDs in the chiral quark soliton model}\label{sec:model}
In this section, we briefly introduce the chiral quark soliton model and express the GTMD correlators as sums over quark states.  We include rotational corrections and extract both the leading and subleading isospin combinations of the quark GTMDs.

\subsection{Model}\label{sec:cqsm}

We use the chiral quark soliton model~\cite{Diakonov:1987ty, Reinhardt:1988fz, Alkofer:1994ph, Christov:1995vm} in the form of Ref.~\cite{Minamiguchi:2026qcb}.  Reference~\cite{Minamiguchi:2026qcb} gives the details of the model.
The quark states are the eigenstates, $\ket{n}$, of the one-particle Hamiltonian,
\begin{equation}
    H_c = - i\gamma^0\gamma^k\partial_k + \gamma^0MU_c^{\gamma_5}\,,
  \label{eq:dirac-spectrum}
\end{equation}
where $M$ is the constituent quark mass and $U_c^{\gamma_5}$ is the self-consistent hedgehog pion field.  
The eigenvalue of $\ket{n}$ is $E_n$.  The valence state is the discrete state with $-M < E_n < M$ in the sector of grand spin $0$ and positive parity.
The occupied states are the states whose energies, $E_n$, do not exceed the energy of the valence state.  All other states are non-occupied.  The valence contribution is the term of the valence state in a sum over the occupied states, and the sea contribution is the total minus the valence contribution.

The spin and isospin of the proton are obtained by quantizing the collective rotation of the soliton.
The angular velocity, $\Omega$, has components $\Omega^a$ with $a = 1, 2, 3$.  Collective quantization replaces $\Omega^a$ by $J^a/I$, where $J^a$ is the proton spin operator and $I$ is the moment of inertia.  Since $I = O(N_c)$ and $J^a = O(N_c^0)$, the angular velocity is $\Omega = O(N_c^{-1})$.
Terms linear in $\Omega$ are required to determine the subleading isospin combinations of the GTMDs.
We evaluate $I$ using the expression given in Ref.~\cite{Pobylitsa:1998tk}.

The contribution of the negative Dirac sea to a sum over the occupied states is ultraviolet divergent.  From the contribution of the negative Dirac continuum, we subtract the corresponding contribution of free quarks.
We regularize the difference by the double Pauli--Villars subtraction~\cite{Kubota:1999hx, Weigel:1999pc, Minamiguchi:2026qcb}.

\subsection{Leading isospin combinations at $O(\Omega^0)$}\label{sec:omega0}
In the model, we first compute the leading isospin combinations at zeroth order in $\Omega$, where the matrix element of the bilocal operator of Eq.~\eqref{eq:correlator} between proton states is~\cite{Diakonov:1996sr, Petrov:1998kf, Ossmann:2004bp}
\begin{align}
    &\mel**{P', S'}{\bar{\psi}\left(-\frac{z}{2}\right)\Gamma\psi\left(\frac{z}{2}\right)}{P, S} = 2\Mcl N_c\delta_{S'S}
    \label{eq:bilocal} \\
    &\qquad \times \sum_{n\in\occ}e^{-iz^0E_n}\mel**{n}{\gamma^0\Gamma e^{i\hat{\bm{p}}\cdot\bm{z}/2}e^{i\bm{\Delta}\cdot\hat{\bm{X}}}e^{i\hat{\bm{p}}\cdot\bm{z}/2}}{n}\,. \nonumber
\end{align}
Here, $\hat{\bm{X}}$ is the quark position operator, $\hat{\bm{p}}$ is the quark momentum operator, and $\Mcl$ is the classical soliton mass, the static energy of the hedgehog, which equals $M_N$ at leading order in $1/N_c$.
The vector $\bm{z} = (\bm{z}_\perp, - z^0)$ is the spatial part of the separation $z$ with $z^+ = 0$.

In Eq.~\eqref{eq:bilocal}, we have set $\mathcal{W} = 1$ because the model contains no explicit gluon fields.
This prescription is also used in CQSM calculations of TMDs~\cite{Wakamatsu:2009fn}.

In QCD, the GTMD expressions for canonical quark OAM and SOC involve a staple-shaped Wilson line~\cite{Lorce:2011kd, Hatta:2011ku}.  For a staple extending along the light-cone direction to $z^- = +\infty$ or $z^- = -\infty$, the Wilson line reduces to unity in the light-cone gauge, $A^+ = 0$, if $\bm{A}_\perp$ vanishes at the corresponding infinity.
Even with this choice, gluonic interaction phases remain encoded in the gauge-dependent QCD wave functions~\cite{Belitsky:2002sm}.
The resulting similarity of the operator forms therefore does not imply an exact identification of the model and QCD correlators.

The model correlator in Eq.~\eqref{eq:bilocal} has no dependence on whether the staple is future- or past-pointing.  Time-reversal symmetry then requires the GTMDs to be real, and their naive time-reversal-odd ($T$-odd) imaginary parts vanish~\cite{Meissner:2009ww}.
This includes the imaginary part of $F_{1, 2}$, which is related to the Sivers function in the forward limit.
The absence of these $T$-odd effects is a limitation of the model~\cite{Wakamatsu:2009fn}.  Within this prescription, the correlator with $\mathcal{W} = 1$ provides an unambiguous definition of the $T$-even GTMDs of the model.

We substitute Eq.~\eqref{eq:bilocal} into Eq.~\eqref{eq:correlator}.
With $z^+ = 0$, the light-cone component of the separation is $z^- = \sqrt{2}z^0$, and the phase in the Breit frame at large $N_c$ is $k\cdot z = x\Mcl z^0 - \kp\cdot\bm{z}_\perp$.
For $\Gamma = \gamma^+$, the matrix $\gamma^0\Gamma$ is $\Gamma_+ / \sqrt{2}$ with the vertex $\Gamma_+ = 1 + \gamma^0\gamma^3$.
We move $e^{i\bm{\Delta}\cdot\hat{\bm{X}}}$ to the left with $[\hat{X}^a, \hat{p}^b] = i\delta^{ab}$ and use $\Delta^3 = - 2\Mcl\xi$.
The integrals over $z$ then give, for an energy $E$,
\begin{equation}
\begin{split}
    &\int\frac{dz^0}{2\pi}\int\frac{d^2\bm{z}_\perp}{(2\pi)^2}\,e^{iz^0(x\Mcl - E) - i\kp\cdot\bm{z}_\perp} \\
    &\qquad\qquad\qquad\qquad \times \Gamma_+e^{i\hat{\bm{p}}\cdot\bm{z} / 2}e^{i\bm\Delta\cdot\hat{\bm{X}}}e^{i\hat{\bm{p}}\cdot\bm{z} / 2} \\
    =& \Gamma_+e^{i\bm\Delta\cdot\hat{\bm{X}}}\int\frac{dz^0}{2\pi}\,e^{iz^0[(x + \xi)\Mcl - E - \hat{p}^3]} \\
    &\qquad\qquad\qquad\qquad \times \int\frac{d^2\bm{z}_\perp}{(2\pi)^2}\,e^{i\bm{z}_\perp\cdot(\hat{\bm{p}}_\perp - \kp + \Dp / 2)} \\
    =& \Gamma_+e^{i\bm\Delta\cdot\hat{\bm{X}}}\delta^{(3)}\left(\hat{\bm{p}} - \pket(E)\right)\,,
\end{split}
\label{eq:zint}
\end{equation}
with the momentum,
\begin{equation}
    \pket(E) = \left(\kp - \Dp / 2, \ (x + \xi)\Mcl - E\right)\,.
\label{eq:pket}
\end{equation}
We write the last line of Eq.~\eqref{eq:zint} as $\Dop_+(E)$.  For $\Gamma = \gamma^+\gamma_5$, the same steps with the vertex $\Gamma_5 = (1 + \gamma^0\gamma^3)\gamma_5$ give $\Dop_5(E)$:
\begin{equation}
\begin{split}
    \Dop_5(E) &= \Gamma_5e^{i\bm{\Delta}\cdot\hat{\bm{X}}}\delta^{(3)}\left(\hat{\bm{p}} - \pket(E)\right)\,.
\end{split}
\label{eq:Dop}
\end{equation}
The delta function fixes the momentum of the quark taken out of the initial proton.

The isosinglet operator $\bar{\psi}\Gamma\psi$ gives the isospin combination $u + d$.
The isotriplet operator $\bar{\psi}\tau^3\Gamma\psi$ gives the isospin combination $u - d$, where $\tau^a$ ($a = 1, 2, 3$) are the Pauli matrices in isospin space.
For the isotriplet operator, the quantization of the collective rotation replaces $\tau^3$ by $-(\sigma^a)_{S'S}\tau^a/3$ between proton states, where $\tau^a$ acts on the quark states of the static soliton~\cite{Adkins:1983ya, Christov:1995vm}.
Comparing the results for the two operators with Eq.~\eqref{eq:Wdecomp}, we obtain the coefficients given by the static soliton,
\begin{subequations}
\label{eq:om0}
\begin{align}
    \cS^\upd &= \Mcl N_c\sum_{n\in\occ}\mel{n}{\Dop_+(E_n)}{n}\,,
    \label{eq:om0S}\\
    \tcS^\upd &= \Mcl N_c\sum_{n\in\occ}\mel{n}{\Dop_5(E_n)}{n}\,,
    \label{eq:om0tS}\\
    \cV^{\umd, a} &= - \frac{\Mcl N_c}{3}\sum_{n\in\occ}\mel{n}{\tau^a\Dop_+(E_n)}{n}\,,
    \label{eq:om0V}\\
    \tcV^{\umd, a} &= - \frac{\Mcl N_c}{3}\sum_{n\in\occ}\mel{n}{\tau^a\Dop_5(E_n)}{n}\,.
\label{eq:om0tV}
\end{align}
\end{subequations}

\subsection{Subleading isospin combinations at \texorpdfstring{$\mathcal{O}(\Omega^1)$}{O(Omega^1)}}\label{sec:omega1}
The other four coefficients, which give the subleading isospin combinations, first appear at first order in $\Omega$.  We write the rotating soliton as $A(t)U_cA^\dagger(t)$, where $U_c$ is the static hedgehog field and $A(t)$ is the SU(2) matrix of the collective rotation, and the quark field as $\psi = A(t)\psi'$.
In this rotating frame, the one-particle Hamiltonian is $H_c + \Omega^a\tau^a/2$ with $-iA^\dagger\dot{A} = \Omega^a\tau^a/2$.
In addition, the two quark fields of Eq.~\eqref{eq:bilocal} are at times separated by $z^0$, and the rotation of the soliton between the two times gives the factor $A^\dagger(-z^0/2)A(z^0/2) = 1 + iz^0\Omega^a\tau^a/2$.
After the collective quantization~\cite{Wakamatsu:1990ud}, the two effects together give terms of the form~\cite{Pobylitsa:1998tk, Ossmann:2004bp}
\begin{align}
    &\cA[\hat{Q}_1, \hat{Q}_2] = \sum_{n\in\occ}\sum_{m:E_m\neq E_n}\left[\frac{1}{E_n - E_m} + \frac{1}{2\Mcl}\pdv{x}\right] \nonumber \\
    &\qquad\times\Bigl(\mel{n}{\hat{Q}_1}{m}\mel{m}{\hat{Q}_2(E_n)}{n} \nonumber \\
    &\qquad\qquad\qquad + \mel{n}{\hat{Q}_2(E_n)}{m}\mel{m}{\hat{Q}_1}{n}\Bigr)\,.
    \label{eq:Sdef}
\end{align}
Here $\hat Q_1$ is a matrix in isospin space, and $\hat Q_2(E)$ is an operator that depends on the energy argument $E$.  The sum over $m$ runs over all states with $E_m\neq E_n$, and these states include occupied and non-occupied states.  The first term in the square brackets multiplies the expression in parentheses, and the second term differentiates that expression.
The first term comes from the change of the occupied states and of their time evolution.  The derivative comes from the factor $iz^0$ of the rotation during the time $z^0$, because $iz^0e^{iz^0x\Mcl} = \Mcl^{-1}\partial e^{iz^0x\Mcl} / \partial x$.
In a complete basis, the two orders of $\hat{Q}_1$ and $\hat{Q}_2$ give the same sum in the derivative term.
For the derivative term, Eq.~\eqref{eq:Sdef} uses the average of the two orders.
The coefficients that require the rotation are
\begin{subequations}
\label{eq:om1}
\begin{align}
    \cV^{\upd, a} &= \frac{\Mcl N_c}{4I}\cA\left[\tau^a, \Dop_+\right]\,,
    \label{eq:om1V} \\
    \tcV^{\upd, a} &= \frac{\Mcl N_c}{4I}\cA\left[\tau^a, \Dop_5\right]\,,
    \label{eq:om1tV} \\
    \cS^\umd &= - \frac{\Mcl N_c}{12I}\sum_{b = 1}^{3}\cA\left[\tau^b, \tau^b\Dop_+\right]\,,
    \label{eq:om1S} \\
    \tcS^\umd &= - \frac{\Mcl N_c}{12I}\sum_{b = 1}^{3}\cA\left[\tau^b, \tau^b\Dop_5\right]\,.
    \label{eq:om1tS}
\end{align}
\end{subequations}
At first order in $\Omega$, the same quantization replaces the product of $\tau^3$ and $\Omega^b$ by $-\delta^{ab}\tau^a/(6I)$ plus a spin-dependent term between proton states.

\subsection{Extraction of the individual GTMDs}\label{sec:extract}

We choose the coordinate axes such that $\Dp$ points along the first axis, and write $\kp = |\kp|(\cos\phi, \sin\phi)$.  With these axes, the leading terms of Eq.~\eqref{eq:coefficients} give
\begin{subequations}
\label{eq:extract}
\begin{align}
    F^\upmd_{1, 2} &= - \frac{i\Mcl\cV^{\upmd, 1}}{|\kp|\sin\phi}\,, \\
    F^\upmd_{1, 3} &= \frac{i\Mcl\cV^{\upmd, 2} - |\kp|\cos\phi F^\upmd_{1, 2}}{|\Dp|}\,, \\
    F^\upmd_{1, 4} &= \frac{i\Mcl^2\cV^{\upmd, 3}}{|\kp||\Dp|\sin\phi}\,, \\
    G^\upmd_{1, 2} &= \frac{\Mcl\tcV^{\upmd, 2}}{|\kp|\sin\phi}\,, \\
    G^\upmd_{1, 3} &= \frac{\Mcl\tcV^{\upmd, 1} - |\kp|\cos\phi G^\upmd_{1, 2}}{|\Dp|}\,, \\
    G^\upmd_{1, 4} &= \tcV^{\upmd, 3}\,,
\end{align}
\end{subequations}
for both isospin combinations, and
\begin{subequations}
\label{eq:extract11}
\begin{align}
    F^\upd_{1, 1} &= \cS^\upd\,, \\
    F^\umd_{1, 1} &= \cS^\umd + \frac{i|\Dp|}{2\Mcl}\cV^{\umd, 2}\,, \\
    G^\upd_{1, 1} &= - \frac{i\Mcl^2\tcS^\upd}{|\kp||\Dp|\sin\phi}\,, \\
    G^\umd_{1, 1} &= - \frac{i\Mcl^2\tcS^\umd}{|\kp||\Dp|\sin\phi} + \frac{1}{2}G^\umd_{1, 2}\,.
\end{align}
\end{subequations}
The extraction of $F^\upmd_{1, 2}$ and $G^\upmd_{1, 2}$ requires $|\kp|\sin\phi \neq 0$.  The extraction of $F^\upmd_{1, 3}$, $F^\upmd_{1, 4}$, $G^\upmd_{1, 3}$, and $G^\upmd_{1, 1}$ requires in addition $|\Dp| \neq 0$.  Details of the numerical calculations are given in Appendices~\ref{app:evaluation} and \ref{app:settings}.

\section{Numerical results of GTMDs}\label{sec:results-gtmd}

\begin{figure*}
\includegraphics[width=\textwidth]{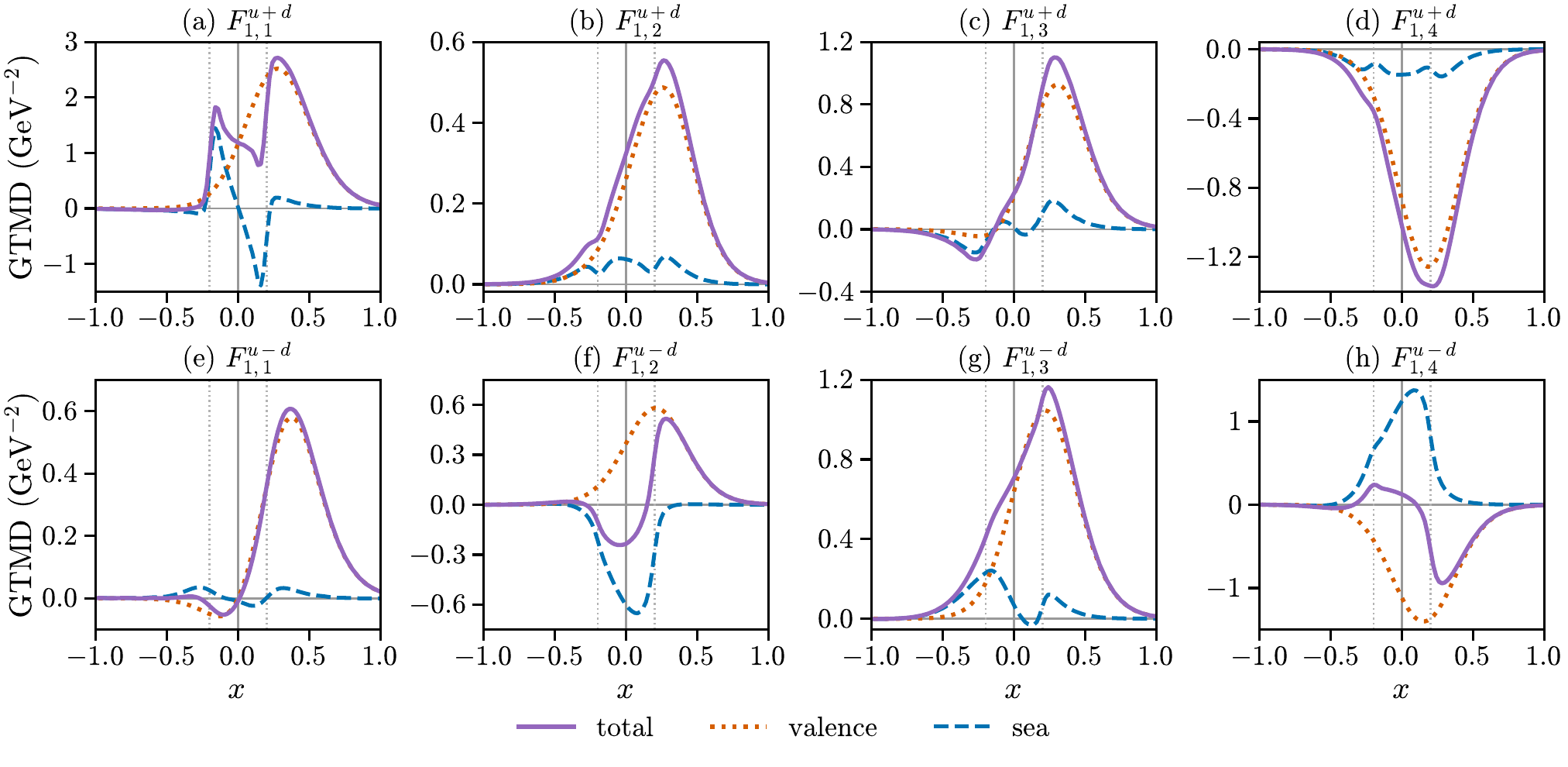}
\caption{Valence contribution (dotted), sea contribution (dashed), and their sum (solid) for $F^\upd_{1, n}$ [(a)--(d)] and $F^\umd_{1, n}$ [(e)--(h)] ($n = 1, \dots, 4$) at $\xi=0.20$, $t=-0.30$~GeV$^2$, $|\kp|=0.3$~GeV, and $\phi=\pi/2$.  The leading isospin combinations are shown in (a), (f), (g), and (h).  The thin gray vertical lines mark $x=\pm\xi$.}
\label{fig:F}
\end{figure*}

\begin{figure*}
\includegraphics[width=\textwidth]{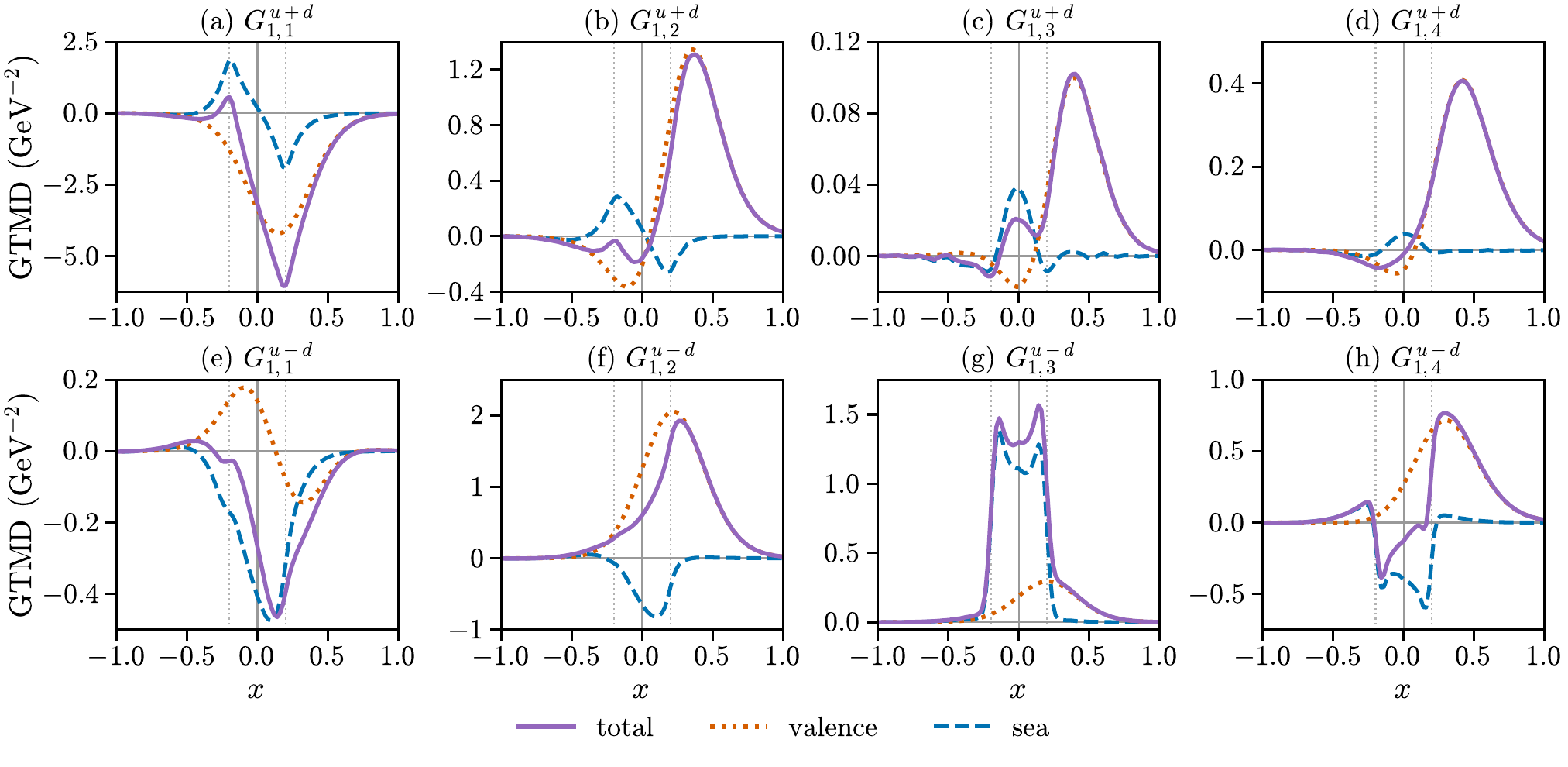}
\caption{Valence contribution (dotted), sea contribution (dashed), and their sum (solid) for $G^\upd_{1, n}$ [(a)--(d)] and $G^\umd_{1, n}$ [(e)--(h)] ($n = 1, \dots, 4$) at the same kinematic point as in Fig.~\ref{fig:F}.  The leading isospin combinations are shown in (a), (f), (g), and (h).  The thin gray vertical lines mark $x=\pm\xi$.}
\label{fig:G}
\end{figure*}

We compute the unpolarized GTMDs, $F_{1, n}$, and the polarized GTMDs, $G_{1, n}$ ($n = 1, \dots, 4$), each for $u + d$ and for $u - d$, at the kinematic point $\xi = 0.20$, $t = - 0.30$~GeV$^2$, $|\kp| = 0.30$~GeV, and $\phi = \pi / 2$, i.e., the vectors, $\kp$ and $\Dp$, are perpendicular.  The values of $\xi$ and $t$ are those of the kinematic points of Ref.~\cite{Minamiguchi:2026qcb}.  The results are shown in Figs.~\ref{fig:F} and \ref{fig:G}.  

We emphasize that the sea quark contributions in Figs.~\ref{fig:F} and \ref{fig:G} are sizable and substantially modify the $x$ dependence of several GTMDs.  This is particularly evident in $G^{\umd}_{1, 3}$, $F^{\umd}_{1, 4}$, and $F^{\upd}_{1, 1}$.

For $G^\umd_{1, 3}$, the sea contribution dominates throughout most of the ERBL region, $|x| < \xi$, and produces the pronounced enhancement shown in Fig.~\ref{fig:G}.  The valence contribution is comparatively small in this region.  At leading order in $1 / N_c$, the relation between the GTMDs and the isotriplet polarized GPD, $\widetilde{E}^\umd$, reads~\cite{Meissner:2009ww}
\begin{equation}
\begin{split}
    \widetilde{E}^\umd(x, \xi, t) =& \frac{2}{\xi}\int d^2\kp\,\left(\frac{\kp\cdot\Dp}{|\Dp|^2}G^\umd_{1, 2} + G^\umd_{1, 3}\right)\,.
\end{split}
\label{eq:EtG}
\end{equation}
In the CQSM, the pion-pole contribution to $\widetilde{E}^\umd$ originates from the long-range pion field of the soliton and is contained in the Dirac-continuum contribution~\cite{Penttinen:1999th}.  This contribution has support only in the ERBL region, and its $x$ dependence is governed by the pion distribution amplitude~\cite{Penttinen:1999th}.  These results suggest that pion exchange plays an important role in the sea-dominated enhancement of $G^\umd_{1, 3}$.

For $F^\umd_{1, 4}$, the sea and valence contributions have opposite signs and comparable magnitudes in the ERBL region, as shown in Fig.~\ref{fig:F}.  The positive sea contribution largely cancels the negative valence contribution and makes the total distribution positive in part of this region.  
We discuss the implications for isotriplet quark OAM in Sec.~\ref{sec:oam-umd}, where we also examine the relation to the recent calculation of Ref.~\cite{Kim:2026gat}.

For $F^\upd_{1, 1}$, the sea contribution is positive near $x = - \xi$ and negative near $x = \xi$ on the ERBL side of the boundaries.
It therefore enhances the distribution near the former boundary and suppresses it near the latter, producing the pronounced structures shown in Fig.~\ref{fig:F}.
At the kinematics shown, the sea contribution is comparable to the valence contribution in the ERBL region and is essential for reproducing these structures. 
A similarly important role of the Dirac sea near the ERBL boundaries was found in the CQSM calculation of the isosinglet GPD, $H^{u+d}$~\cite{Petrov:1998kf}.

\begin{figure*}
\includegraphics[width=\textwidth]{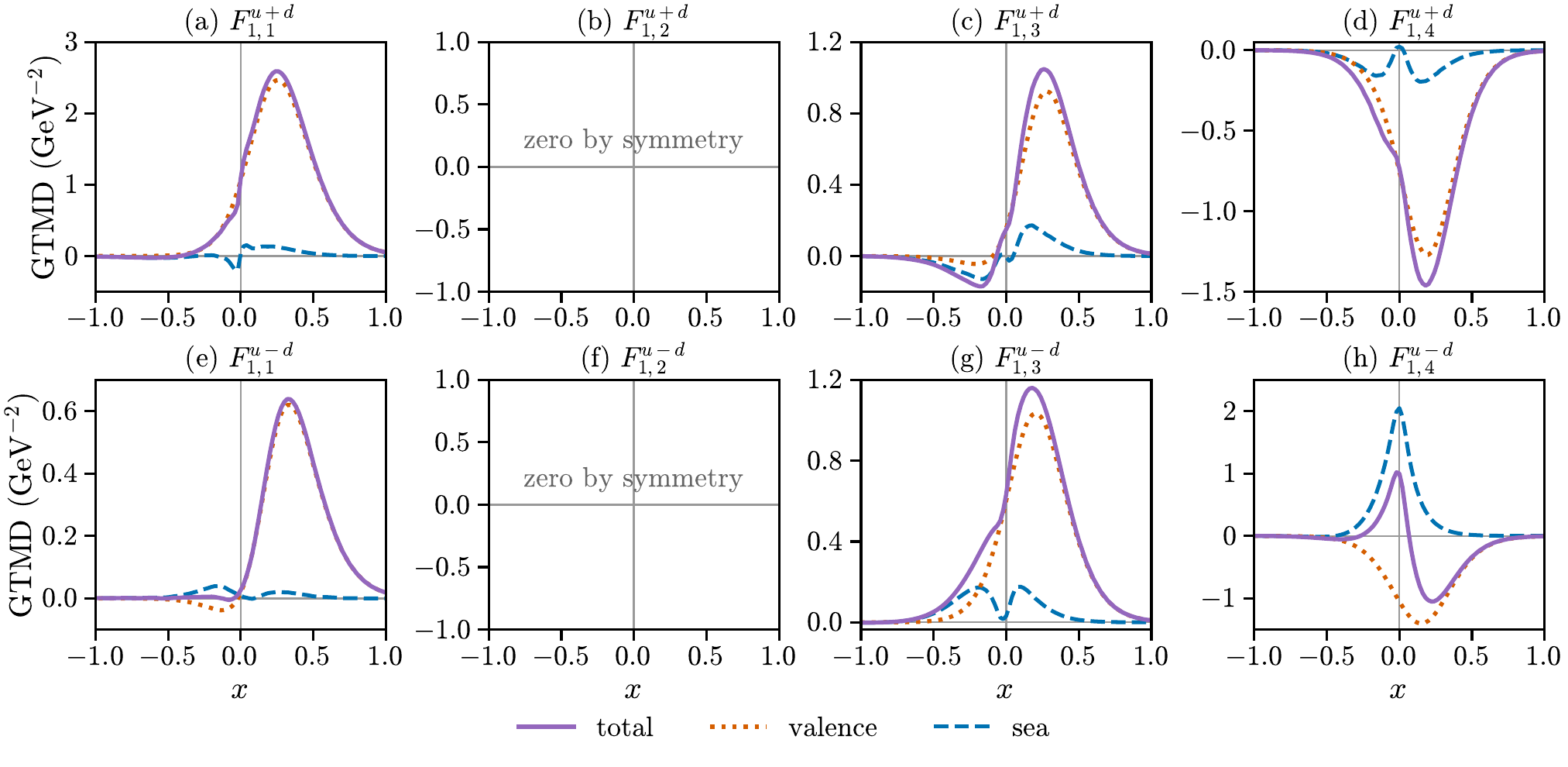}
\caption{Same as Fig.~\ref{fig:F}, but at $\xi = 0$.}
\label{fig:Fxi0pi2}
\end{figure*}

\begin{figure*}
\includegraphics[width=\textwidth]{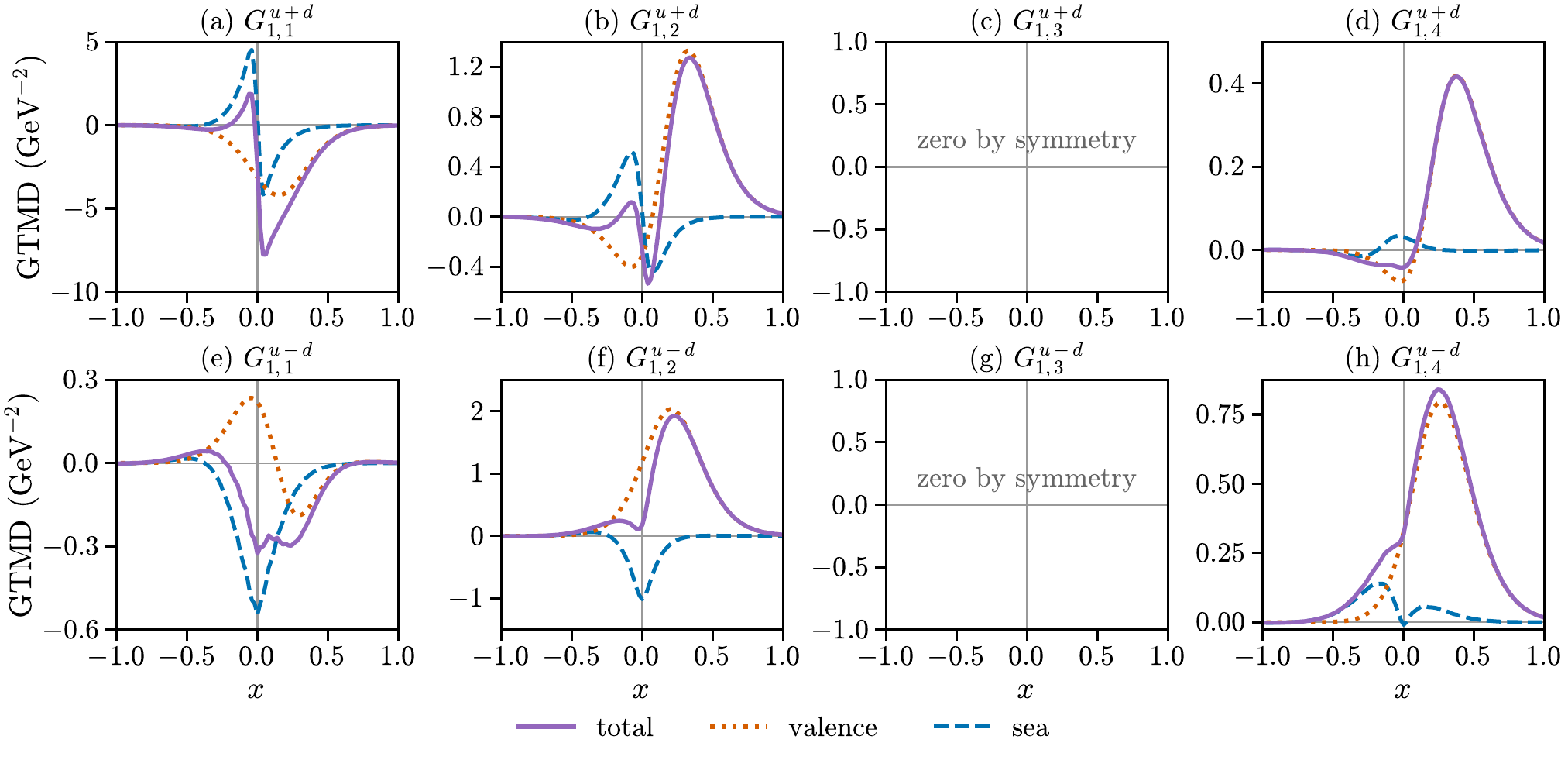}
\caption{Same as Fig.~\ref{fig:G}, but at $\xi=0$.}
\label{fig:Gxi0pi2}
\end{figure*}

\begin{figure*}
\includegraphics[width=\textwidth]{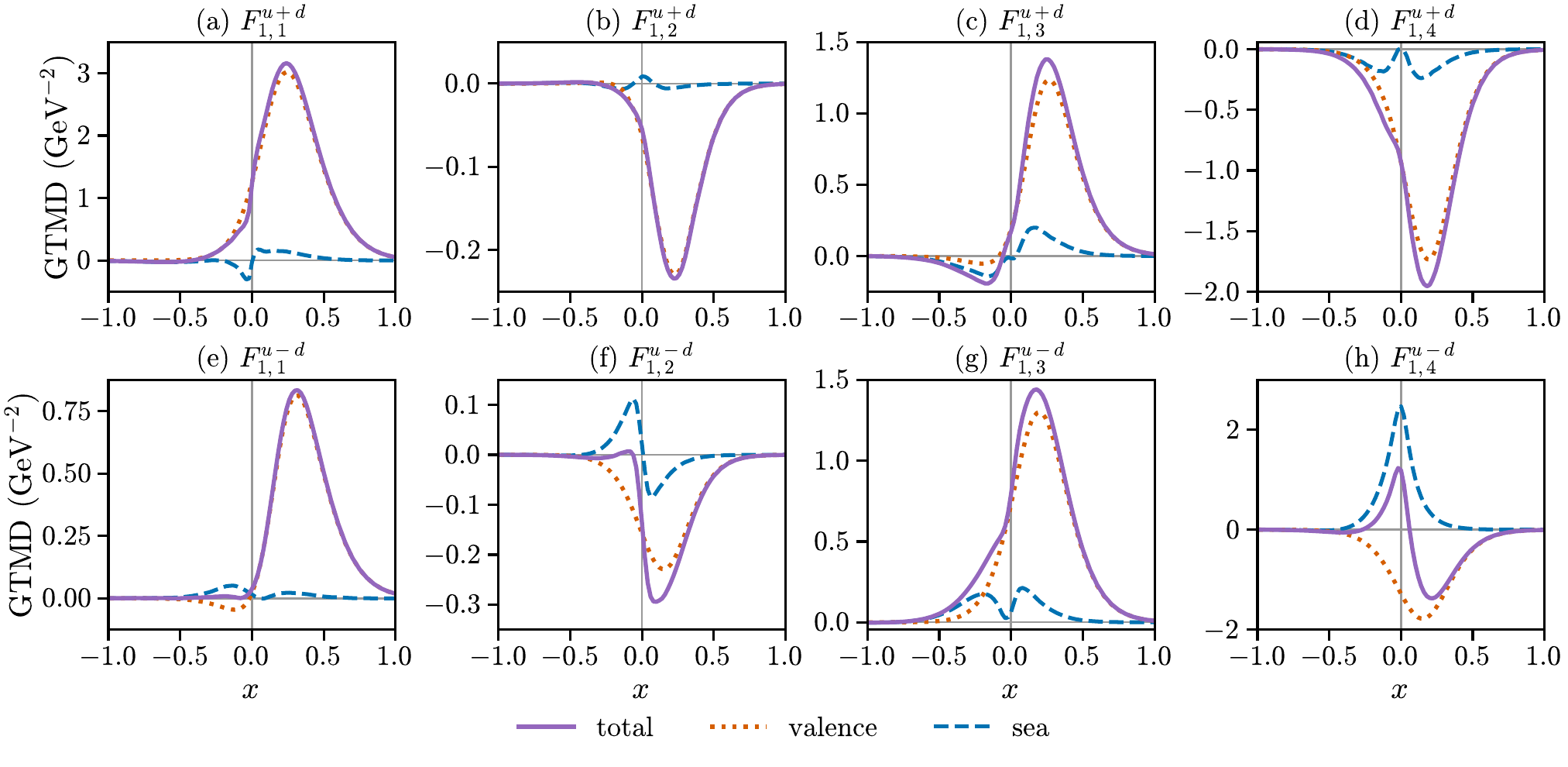}
\caption{Same as Fig.~\ref{fig:Fxi0pi2}, but at $\phi = \pi / 4$.}
\label{fig:Fxi0}
\end{figure*}

\begin{figure*}
\includegraphics[width=\textwidth]{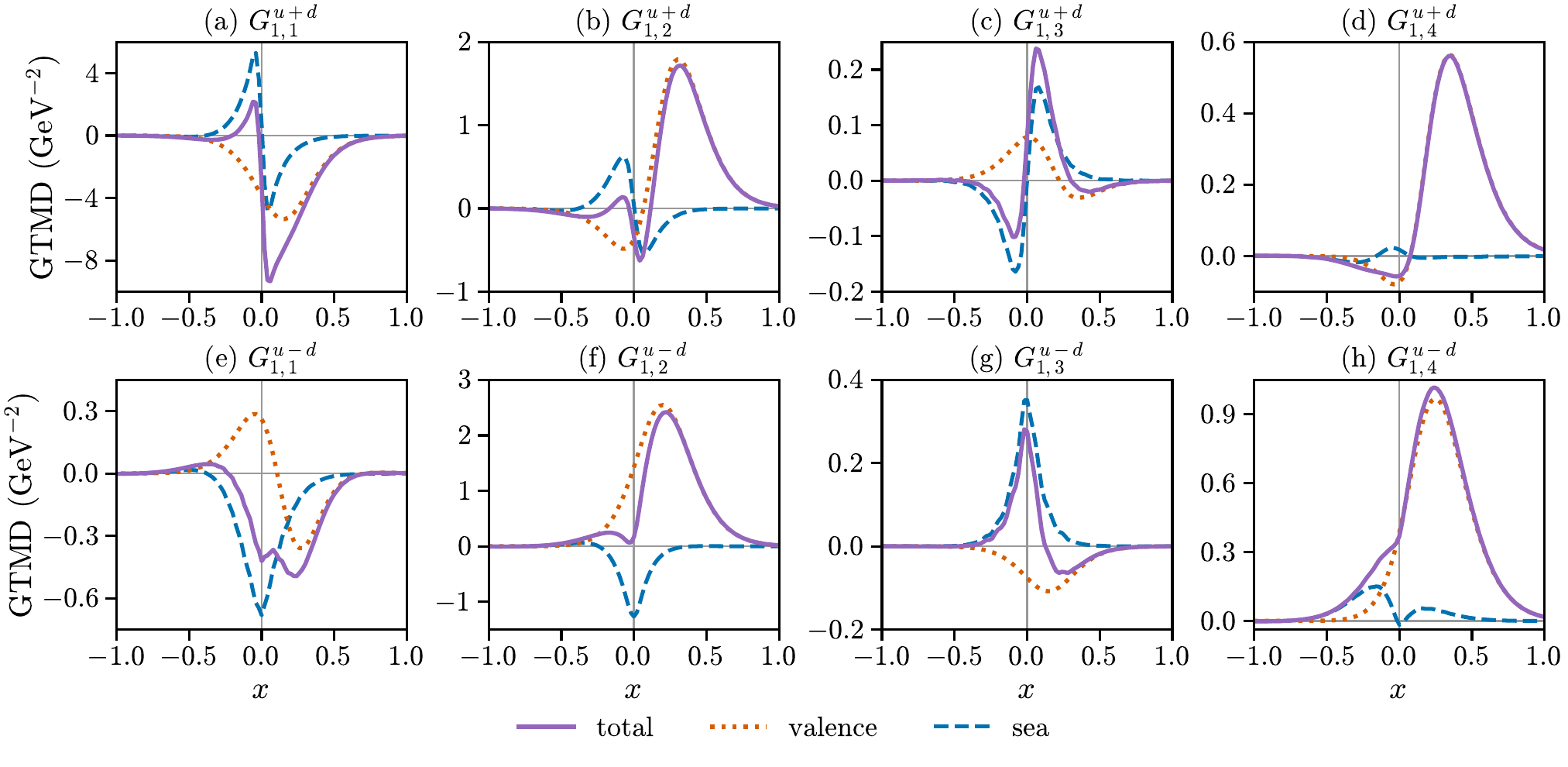}
\caption{Same as Fig.~\ref{fig:Gxi0pi2}, but at $\phi = \pi / 4$.}
\label{fig:Gxi0}
\end{figure*}

We also compute the GTMDs at $\xi = 0$, keeping $t$, $|\kp|$, and $\phi = \pi / 2$ unchanged.  The results are shown in Figs.~\ref{fig:Fxi0pi2} and \ref{fig:Gxi0pi2}.  Hermiticity makes the real parts of $F_{1, 2}$ and $G_{1, 3}$ odd under the simultaneous reversal of $\xi$ and $\kp\cdot\Dp$~\cite{Meissner:2009ww}.  Since the GTMDs are real in this model, both isospin combinations of these two GTMDs must vanish at $\xi = 0$ and $\phi = \pi / 2$, where $\kp\cdot\Dp = 0$.  Our numerical results satisfy this constraint within numerical precision.

We also consider $\phi = \pi / 4$, for which $\kp\cdot\Dp \neq 0$ and these GTMDs are no longer required to vanish.  The results are shown in Figs.~\ref{fig:Fxi0} and \ref{fig:Gxi0}.  At both values of $\phi$, the sea contributions to $F^\upd_{1, 1}$ and $G^\umd_{1, 4}$ are relatively small, in contrast to the pronounced contributions in the ERBL region at $\xi = 0.20$.

\section{Wigner distribution}\label{sec:wigner}

A particularly useful application of GTMDs is the study of quark Wigner distributions, which describe correlations between the transverse position and momentum of quarks at a fixed longitudinal momentum fraction~\cite{Ji:2003ak, Belitsky:2003nz, Lorce:2011kd}.  They provide a five-dimensional phase-space description of the proton.

At zero skewness, the Wigner distributions are defined by the Fourier transform of the GTMD correlators with respect to the transverse momentum transfer~\cite{Lorce:2011kd},
\begin{equation}
\begin{split}
    \rho^{[\Gamma], q}_{S'S}(x, \bm{b}_\perp, \kp) = \int\frac{d^2\Dp}{(2\pi)^2}\,e^{-i\Dp\cdot\bm{b}_\perp}W^{[\Gamma], q}_{S'S}\eval_{\xi = 0}\,.
\end{split}
\label{eq:wigner_general}
\end{equation}
Here, $\Gamma$ specifies the quark polarization.  
The vector $\bm{b}_\perp$ is the transverse position of the quark relative to the transverse center of momentum of the proton.

We focus on the Wigner distribution $\rho_{UU}^q$ of unpolarized quarks in an unpolarized proton.  It is obtained as $\rho_{UU}^q = \frac{1}{2}\sum_{S = \pm 1/2}\rho^{[\gamma^+], q}_{SS}$.
For the isosinglet combination, $\rho^\upd = \rho_{UU}^u + \rho_{UU}^d$, we get~\cite{Lorce:2011kd}
\begin{equation}
    \rho^\upd(x, \bm{b}_\perp, \kp) = \int\frac{d^2\Dp}{(2\pi)^2}\,
    e^{-i\Dp\cdot\bm{b}_\perp}F^\upd_{1,1}\,,
\label{eq:wigner}
\end{equation}
where $F^\upd_{1, 1}$ is evaluated at $\xi = 0$.

Since $F^\upd_{1, 1}$ is real in this model, hermiticity makes it an even function of $\kp\cdot\Dp$ at $\xi = 0$~\cite{Meissner:2009ww}.  We therefore expand it in the angle $\phi$ between $\kp$ and $\Dp$ as
\begin{equation}
    F^\upd_{1, 1} = \sum_{\kappa = 0}^\infty a_\kappa(x, |\kp|, |\Dp|)\cos2\kappa\phi\,.
\label{eq:wignerharm}
\end{equation}
The expansion reduces the two-dimensional integral of Eq.~\eqref{eq:wigner} to one-dimensional integrals over $|\Dp|$.  We write $\theta$ for the angle from $\bm b_\perp$ to $\kp$.  The integral over the direction of $\Dp$ in Eq.~\eqref{eq:wigner} then gives
\begin{equation}
\begin{aligned}
    \rho^\upd &= \sum_{\kappa = 0}^\infty(-1)^\kappa\cos2\kappa\theta \\
    &\qquad\times\int_0^\infty\frac{d|\Dp|\,|\Dp|}{2\pi}J_{2\kappa}\left(|\bm b_\perp||\Dp|\right)a_\kappa\,,
\end{aligned}
\label{eq:wignerbessel}
\end{equation}
where $J_{2\kappa}$ is the Bessel function.
The function, $\bar\rho^\upd$, is the average of $\rho^\upd$ over $\theta$.
Appendix~\ref{app:wigner} gives the numerical evaluation of Eq.~\eqref{eq:wignerbessel}.
 
\begin{figure*}
\includegraphics[width=0.95\textwidth]{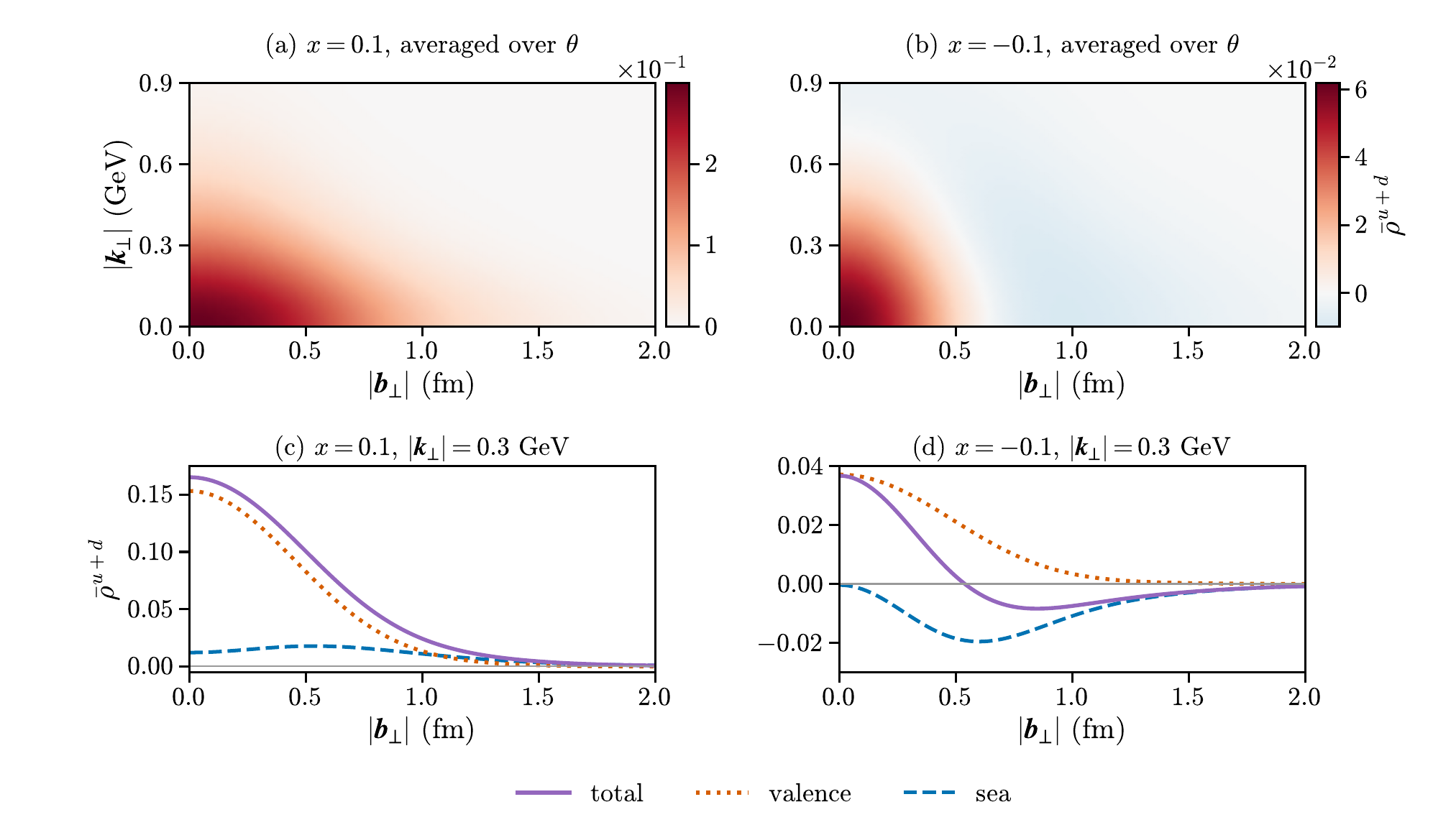}
\caption{Wigner distribution of unpolarized quarks in the unpolarized proton at $x = 0.1$ [(a) and (c)] and at $x = -0.1$ [(b) and (d)].
(a) and (b): $\bar{\rho}^\upd$ as a function of $|\bm{b}_\perp|$ and $|\kp|$.  (c) and (d): $\bar{\rho}^\upd$ at $|\kp| = 0.3$~GeV for the total (solid), the valence contribution (dotted), and the sea contribution (dashed).}
\label{fig:wigner}
\end{figure*}

Figure~\ref{fig:wigner} shows the angular-averaged Wigner distribution $\bar\rho^\upd$ at $x = 0.1$ and $x = -0.1$.
For $x < 0$, $\bar\rho^\upd$ is minus the distribution of the antiquarks with momentum fraction $-x$,
\begin{equation}
    \bar{\rho}^{\bar{u} + \bar{d}}(- x, |\bm{b}_\perp|, |\kp|) = -\bar{\rho}^\upd(x, |\bm b_\perp|, |\kp|)\,.
\label{eq:rhoanti}
\end{equation}

\section{Orbital angular momentum}\label{sec:oam}
The GTMD $F^\upmd_{1, 4}$ gives access to the OAM of the quarks~\cite{Lorce:2011kd, Hatta:2011ku}.

The Wigner distribution of unpolarized quarks in the longitudinally polarized proton is obtained from $F^\upmd_{1, 4}$~\cite{Lorce:2011kd},
\begin{equation}
    \rho^\upmd_{LU} = - \frac{\epsilon^{-+ij}k_\perp^i}{M_N^2}\frac{\partial}{\partial b_\perp^j}\int\frac{d^2\Dp}{(2\pi)^2}\,e^{-i\Dp\cdot\bm{b}_\perp}F^\upmd_{1, 4}\,,
\label{eq:wignerLU}
\end{equation}
where $F^\upmd_{1, 4}$ is evaluated at $\xi = 0$.  The moment of $\rho^\upmd_{LU}$ with $(\bm{b}_\perp\times\kp)_3$, integrated over $x$, $\bm{b}_\perp$, and $\kp$, gives the OAM of the quarks~\cite{Lorce:2011kd},
\begin{equation}
\begin{split}
    L^\upmd &= \int dx\int d^2\bm{b}_\perp\int d^2\kp\,(\bm{b}_\perp\times\kp)_3\rho^\upmd_{LU}\,.
\end{split}
\label{eq:Lq}
\end{equation}
The proton is polarized along the third axis, so rotational symmetry about this axis is preserved.  The Wigner distribution $\rho^\upmd_{LU}$ therefore depends on the directions of $\bm b_\perp$ and $\kp$ only through their relative angle $\theta$.  Using $(\bm b_\perp\times\kp)_3 = |\bm b_\perp||\kp|\sin\theta$, we write the OAM distribution in $x$, $|\bm b_\perp|$, and $|\kp|$ as
\begin{equation}
    \frac{d^3L^\upmd}{dx\,d|\bm b_\perp|\,d|\kp|} = 2\pi|\bm b_\perp||\kp|\int_0^{2\pi}d\theta\,(\bm b_\perp\times\kp)_3\rho^\upmd_{LU}\,.
\label{eq:bxkL}
\end{equation}
We expand $F^\upmd_{1,4}$ in $\cos2\kappa\phi$ in the same way as $F^\upd_{1,1}$ in Eq.~\eqref{eq:wignerharm}.

We note that, in QCD, the GTMD moment with a staple-shaped Wilson line extending to light-cone infinity gives the gauge-invariant extension of the Jaffe--Manohar canonical OAM, whereas a straight Wilson line gives the Ji kinetic OAM~\cite{Hatta:2011ku, Burkardt:2012sd}.
Because the model correlator is defined with $\mathcal{W} = 1$ (Sec.~\ref{sec:omega0}), it does not distinguish between these two paths.
The resulting OAM operator contains ordinary derivatives and defines the canonical OAM of the effective quark fields.

For the flavor-nonsinglet combination, the kinetic OAM is represented in the chiral effective theory from the instanton vacuum by a different operator, which also contains an interaction with the chiral field~\cite{Kim:2026gat}.  We return to this point in Sec.~\ref{sec:oam-umd}.

\subsection{Isosinglet orbital angular momentum}\label{sec:oam-upd}

\begin{figure*}
\includegraphics[width=0.95\textwidth]{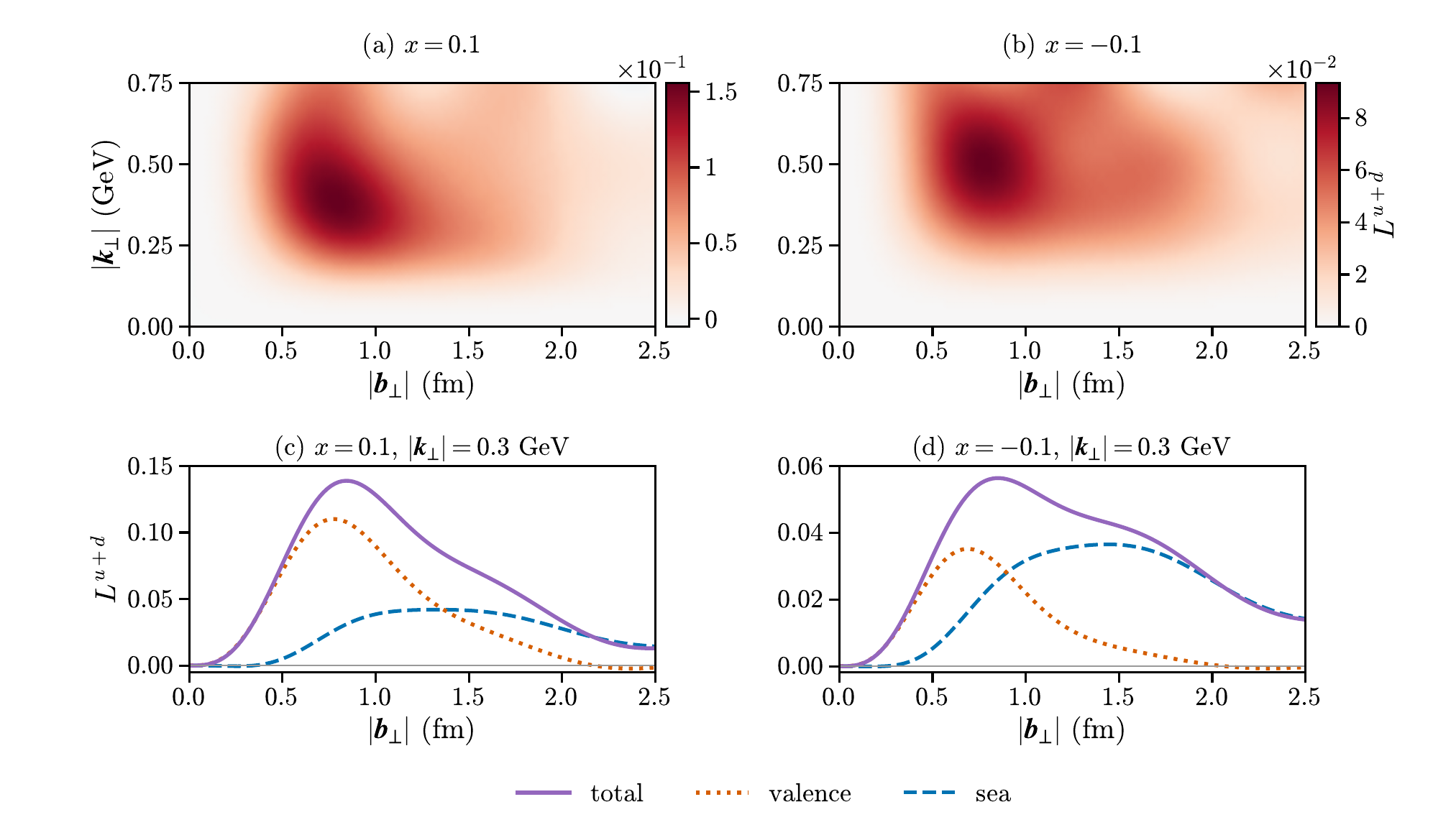}
\caption{Distribution of the isosinglet OAM, $d^3L^\upd/(dx\,d|\bm b_\perp|\,d|\kp|)$ in Eq.~\eqref{eq:bxkL}, at $x = 0.1$ [(a) and (c)] and at $x = -0.1$ [(b) and (d)].  (a) and (b): OAM distribution as a function of $|\bm b_\perp|$ and $|\kp|$ for $|\kp| \le 0.75$~GeV (Appendix~\ref{app:wigner}).  (c) and (d): OAM distribution at $|\kp| = 0.3$~GeV for the total (solid), the valence contribution (dotted), and the sea contribution (dashed).}
\label{fig:wignerUL_upd}
\end{figure*}

Figure~\ref{fig:wignerUL_upd} shows the isosinglet OAM distribution in Eq.~\eqref{eq:bxkL} at $x = 0.1$ and at $x = -0.1$.

At $x = 0.1$, the distribution is positive in Fig.~\ref{fig:wignerUL_upd}(a).
At $|\kp| = 0.3$~GeV, its valence and sea contributions are both positive in almost the whole range of Fig.~\ref{fig:wignerUL_upd}(c).
Quarks with momentum fraction $0.1$ thus have OAM parallel to the spin of the proton.

At $x = -0.1$, the distribution is also positive in the whole region of Fig.~\ref{fig:wignerUL_upd}(b).
Since the distribution at $x < 0$ is the distribution of the antiquarks with momentum fraction $-x$, the antiquarks with momentum fraction $0.1$ also have OAM parallel to the spin of the proton.

The OAM distribution in Fig.~\ref{fig:wignerUL_upd} vanishes at $|\bm b_\perp| = 0$ and reaches its maximum at a finite transverse distance.  The vanishing at the transverse center follows from the factor $(\bm b_\perp\times\kp)_3$.  After averaging over the direction of $\kp$, the corresponding density in the transverse position plane exhibits an annular profile as in Ref.~\cite{Fukushima:2026wwc}.

The sea contribution in Fig.~\ref{fig:wignerUL_upd} extends to much larger transverse distances than the valence contribution and dominates the OAM distribution in the peripheral region.  This extended tail is consistent with the pion cloud surrounding the soliton core.  The spatial extent of the pion cloud is expected to be enhanced in the chiral limit, $m_\pi = 0$, used in the present calculation, where the asymptotic pion field lacks the exponential suppression associated with a finite pion mass.

In the present model, which has no explicit gluon degrees of freedom, the proton spin is carried entirely by the quark and antiquark fields~\cite{Wakamatsu:1999nj, Wakamatsu:2005vk, Wakamatsu:2010nsh}.  The isosinglet spin sum rule therefore reads
\begin{equation}
    L^\upd + S^\upd = \frac{1}{2}\,,
\label{eq:model_spin_sum}
\end{equation}
where $S^\upd = \Delta\Sigma/2$ and $\Delta\Sigma = \int dx\,\widetilde H^\upd(x, 0, 0)$.

Evaluating Eq.~\eqref{eq:Lq} as described in Appendix~\ref{app:settings}, we obtain $L^\upd = 0.314$, with valence and sea contributions of $0.123$ and $0.191$, respectively.
Independently, we calculate the helicity distribution in the same model using the method of Ref.~\cite{Minamiguchi:2026uaq} and obtain $\Delta\Sigma = 0.373$, corresponding to $S^\upd = 0.186$.
These independently calculated spin and orbital contributions satisfy the proton spin sum rule, Eq.~\eqref{eq:model_spin_sum}, within the numerical uncertainties.

\subsection{Isotriplet orbital angular momentum}\label{sec:oam-umd}

\begin{figure*}
\includegraphics[width=0.95\textwidth]{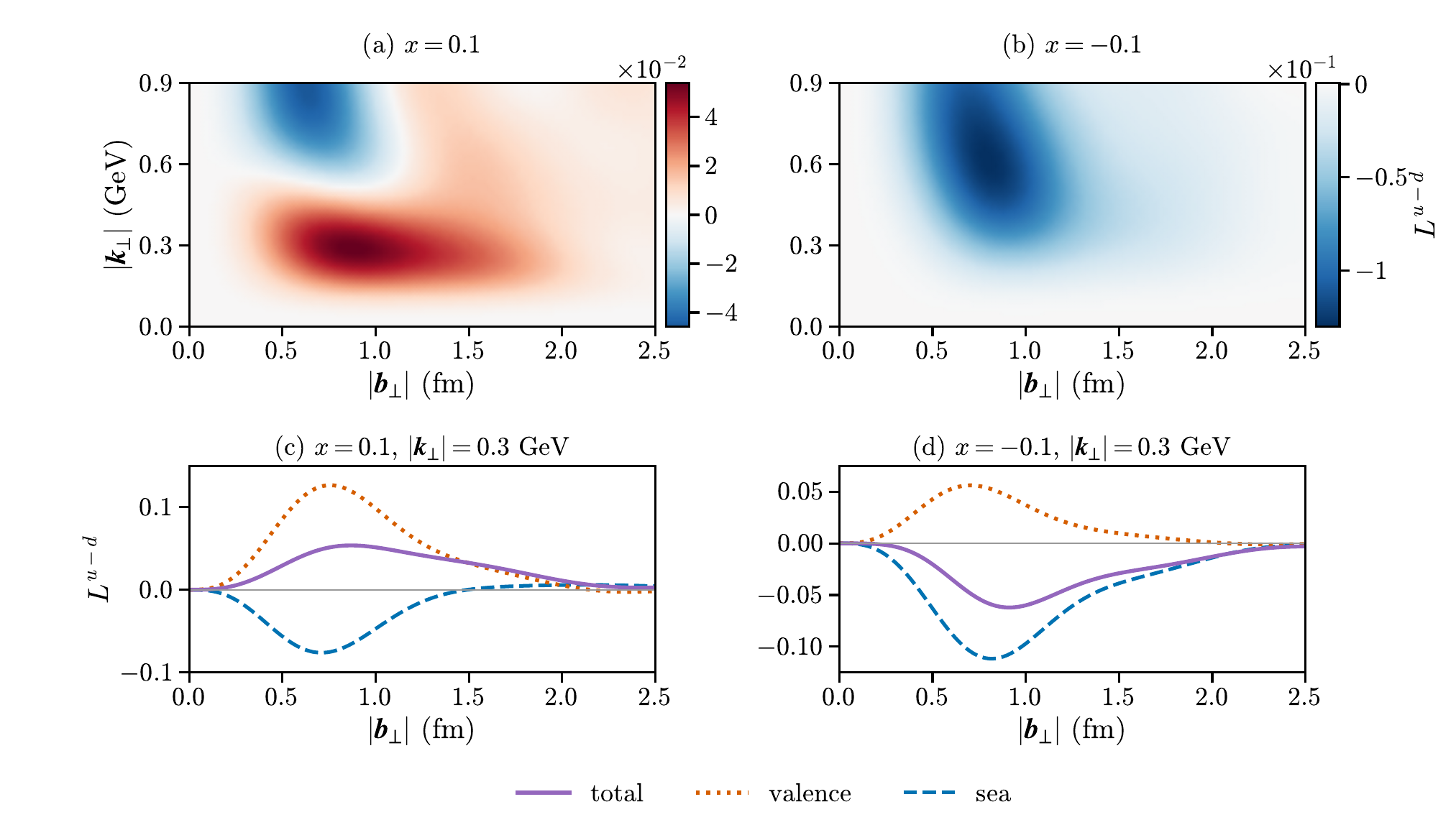}
\caption{Distribution of the isotriplet OAM, $d^3L^\umd/(dx\,d|\bm b_\perp|\,d|\kp|)$ in Eq.~\eqref{eq:bxkL}, at $x = 0.1$ [(a) and (c)] and at $x = -0.1$ [(b) and (d)].  (a) and (b): OAM distribution as a function of $|\bm b_\perp|$ and $|\kp|$.  (c) and (d): OAM distribution at $|\kp| = 0.3$~GeV for the total (solid), the valence contribution (dotted), and the sea contribution (dashed).}
\label{fig:wignerUL_umd}
\end{figure*}

Figure~\ref{fig:wignerUL_umd} shows the isotriplet OAM distribution in Eq.~\eqref{eq:bxkL} at $x = 0.1$ and at $x = -0.1$.
For comparison to Ref.~\cite{Kim:2026gat}, we consider the kinetic OAM, $L^\umd_{\mathrm{kin}} = J^\umd - S^\umd$, where $J^\umd$ and $S^\umd$ are the isotriplet combinations of the total quark angular momentum and quark spin, respectively.
They are given by moments of the unpolarized GPDs and the helicity distribution~\cite{Ji:1996ek},
\begin{align}
    J^\umd &= \frac{1}{2}\int dx\,x\left[H^\umd(x,0,0) + E^\umd(x,0,0)\right]\,,
\label{eq:Jumd}\\
    S^\umd &= \frac12\int dx\,\widetilde H^\umd(x,0,0) = \frac{g_A}{2}\,.
\label{eq:Sumd}
\end{align}
Here, $\widetilde H^\umd(x,0,0)$ is the isotriplet helicity PDF, whose integral gives the axial charge $g_A$.
At leading order in $1/N_c$, we retain only the contribution of $E^\umd$ to $J^\umd$.
We calculate $E^\umd$ using the method of Ref.~\cite{Minamiguchi:2026qcb} and evaluate $g_A$ as a sum over the occupied states of the static soliton.
We obtain $J^\umd = 0.223$ and $g_A = 1.049$, giving $L^\umd_{\mathrm{kin}} = -0.301$.

The OAM, $L^\umd$, obtained from the GTMDs in Eq.~\eqref{eq:Lq} and the kinetic OAM, $L^\umd_{\mathrm{kin}}$, are defined by different operators and need not coincide.
Their relation is explained by the effective-operator analysis of Ref.~\cite{Kim:2023pll, Kim:2026gat}.
When the QCD kinetic OAM operator is matched to the chiral effective theory, the gauge field in the covariant derivative generates an additional component, $L^\umd_{\mathrm{pot}}$, with the chiral field.
Then, we have
\begin{equation}
    L^\umd_{\mathrm{kin}} = L^\umd + L^\umd_{\mathrm{pot}}\,.
\label{eq:oam_kin_pot}
\end{equation}
Here, $L^\umd$ is the contribution from ordinary derivatives, which is evaluated through the GTMD moment in our calculation.

We independently evaluate the potential contribution, $L^\umd_{\mathrm{pot}}$, in Ref.~\cite{Kim:2026gat} in the prescription adopted in this paper and obtain $L^\umd_{\mathrm{pot}} = -0.193$.  This agrees with the difference $L^\umd_{\mathrm{kin}} - L^\umd = -0.196$ within the numerical uncertainties.

\section{Spin--orbit correlation}\label{sec:soc}
GTMDs also provide access to the quark SOC, which characterizes the relative alignment of quark spin and OAM in the proton~\cite{Lorce:2014mxa}.

The Wigner distribution of longitudinally polarized quarks in the unpolarized proton is obtained from $G_{1,1}$~\cite{Lorce:2011kd},
\begin{equation}
    \rho^\upd_{UL} = \frac{\epsilon^{-+ij}k_\perp^i}{M_N^2}\frac{\partial}{\partial b_\perp^j}\int\frac{d^2\Dp}{(2\pi)^2}\,e^{-i\Dp\cdot\bm{b}_\perp}G^\upd_{1, 1}\,,
\label{eq:wignerUL}
\end{equation}
where $G^\upd_{1, 1}$ is evaluated at $\xi = 0$.  The derivative appears because $G_{1, 1}$ enters the correlator~\eqref{eq:defG} with the factor $\epsilon^{-+ij}k_\perp^i\Delta_\perp^j$.  The moment of $\rho^\upd_{UL}$ with $(\bm{b}_\perp\times\kp)_3$ gives the SOC of the quarks~\cite{Lorce:2014mxa},
\begin{equation}
\begin{split}
    C^\upd_z &= \int dx\int d^2\bm{b}_\perp\int d^2\kp\,(\bm{b}_\perp\times\kp)_3\rho^\upd_{UL}\,.
\end{split}
\label{eq:Cz}
\end{equation}
As in the OAM case, the distribution of $C^\upd_z$ in $x$, $|\bm b_\perp|$, and $|\kp|$ is
\begin{equation}
    \frac{d^3C^\upd_z}{dx\,d|\bm b_\perp|\,d|\kp|} = 2\pi|\bm{b}_\perp||\kp|\int_0^{2\pi}d\theta\,(\bm{b}_\perp\times\kp)_3\rho^\upd_{UL}\,.
\label{eq:bxk}
\end{equation}
We expand $G^\upd_{1, 1}$ in $\cos2\kappa\phi$ in the same way as $F^\upd_{1, 1}$ in Eq.~\eqref{eq:wignerharm}.

\begin{figure*}
\includegraphics[width=0.95\textwidth]{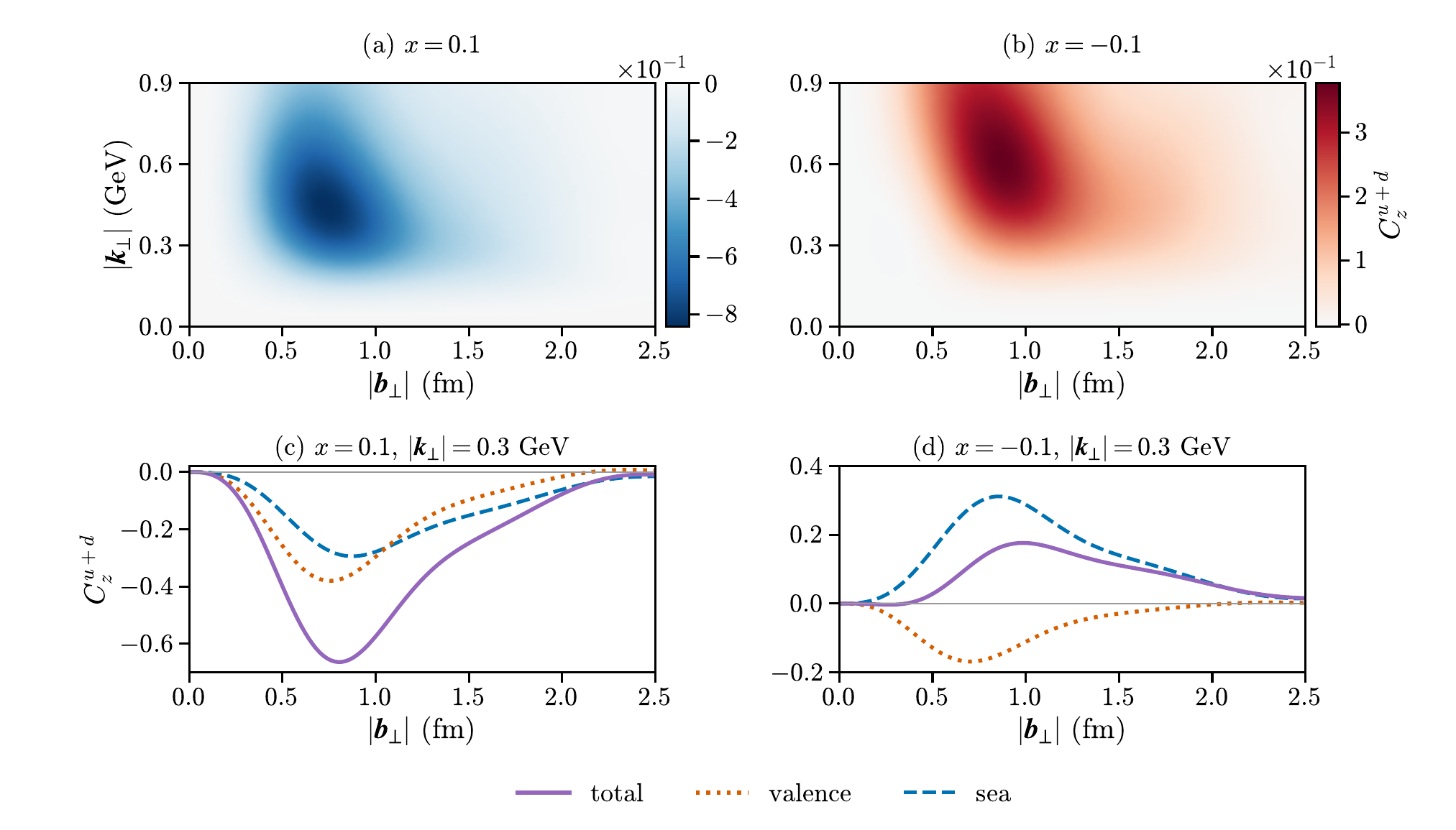}
\caption{Distribution of the SOC, $d^3C^\upd_z/(dx\,d|\bm b_\perp|\,d|\kp|)$ in Eq.~\eqref{eq:bxk}, at $x = 0.1$ [(a) and (c)] and at $x = - 0.1$ [(b) and (d)].  (a) and (b): SOC as a function of $|\bm b_\perp|$ and $|\kp|$.   (c) and (d): SOC at $|\kp| = 0.3$~GeV for the total (solid), the valence contribution (dotted), and the sea contribution (dashed).}
\label{fig:wignerUL}
\end{figure*}

Figure~\ref{fig:wignerUL} shows the SOC distribution in Eq.~\eqref{eq:bxk} at $x = 0.1$ and at $x = -0.1$.  The distribution at $x < 0$ is minus the distribution of the antiquarks with momentum fraction $-x$.

We find $C^\upd_z = -0.43$, with a valence contribution of $-0.42$ and a net sea contribution of approximately $-0.01$.  The quark and antiquark SOCs are $-0.83$ and $-0.40$, respectively.  Both at $x = \pm 0.1$ and after the integration over $x$, the quarks and the antiquarks thus have spin anti-aligned with their OAM.

The integrated SOC involves the difference between the quark and antiquark SOCs~\cite{Lorce:2014mxa, Hatta:2024otc},
\begin{equation}
\begin{split}
    C^\upd_z &= \int_0^1dx\,\left[C^\upd_z(x) - C_z^{\bar{u} + \bar{d}}(x)\right]\,,
\end{split}
\label{eq:Cz_quark_antiquark}
\end{equation}
where $C^\upd_z(x)$ denotes the $x$-dependent SOC obtained by integrating Eq.~\eqref{eq:bxk} over the transverse variables, and the antiquark distribution satisfies $C_z^{\bar u+\bar d}(x) = -C^\upd_z(-x)$ for $x > 0$.  The small net sea contribution is consistent with a near cancellation between the sea-quark and sea-antiquark SOCs, which enter Eq.~\eqref{eq:Cz_quark_antiquark} with opposite signs.  The integrated SOC is thus nearly saturated by the valence contribution, despite the sizable sea contributions at fixed $x$.

\section{Conclusions}\label{sec:conclusions}

We calculated the vector and axial-vector quark GTMDs in the chiral quark soliton model and used them to study Wigner distributions, OAM, and SOC.  By explicitly including the regularized Dirac sea, the calculation extends earlier CQSM studies restricted to the three-quark sector and resolves the contribution of the nonperturbative sea to quark phase-space correlations.

A central result is the determination of the OAM and SOC distributions in the $(|\bm b_\perp|, |\kp|)$ plane at fixed $x$.
These distributions resolve the transverse structure of the OAM and SOC separately for quarks and antiquarks.
At the momentum fractions studied, both quarks and antiquarks exhibit isosinglet OAM aligned with the proton spin and an anti-alignment between their own spin and OAM.  The sea accounts for approximately $60\%$ of the integrated isosinglet OAM, whereas sizable sea-quark and sea-antiquark contributions nearly cancel in the integrated isosinglet SOC.  The phase-space distributions thus reveal the structure underlying these integrated quantities.

Lattice calculations of GTMD moments and proposed measurements in exclusive processes offer opportunities to test the orbital structure of the proton.
The present results provide a nonperturbative picture of how quark and antiquark OAM correlate with the proton spin and with their own spin.  An important next step is to determine how the predicted transverse structure, alignment patterns, and quark--antiquark cancellations enter accessible observables.  Such studies would help clarify the role of the nonperturbative sea in quark OAM and SOC.

\begin{acknowledgments}
This work was carried out during the long-term workshop ``New Frontiers in QCD 2026'' at the Yukawa Institute for Theoretical Physics (YITP-T-26-01).
Discussions during the workshop were useful to complete this work.  The authors used Anthropic's Claude and OpenAI's ChatGPT to assist with the implementation and debugging of the numerical code and in the preparation of the manuscript.  This work was partially supported by IIW, WINGS Program, the University of Tokyo (J.M.) and by FoPM, WINGS Program, the University of Tokyo (T.U.).
\end{acknowledgments}

\appendix

\section{Numerical procedure}\label{app:evaluation}

We evaluate Eqs.~\eqref{eq:om0} and \eqref{eq:om1} with the method of Ref.~\cite{Minamiguchi:2026qcb} in the Kahana--Ripka basis~\cite{Kahana:1984be} in a spherical box of radius $D$.  In this basis, the magnitude $|\bm{p}|$ of the quark momentum takes discrete values spaced by about $\pi / D$.  We therefore replace the delta function in Eq.~\eqref{eq:Dop} by the normalized Gaussian~\cite{Diakonov:1997vc, Minamiguchi:2026qcb}
\begin{equation}
\begin{aligned}
    &\delta^{(3)}\left(\hat{\bm{p}} - \pket(E)\right) \\
    &\qquad\to\frac{1}{(\gamma\Mcl\sqrt\pi)^3}\exp[-\frac{|\hat{\bm{p}} - \pket(E)|^2}{(\gamma\Mcl)^2}]\,,
\end{aligned}
\label{eq:Gauss}
\end{equation}
where $\gamma$ is a dimensionless parameter.  The width $\gamma\Mcl$ must not be smaller than $\pi / D$, because a narrower Gaussian leaves a separate peak at each discrete value of $|\bm{p}|$.

We expand in multipoles both the factor $e^{i\bm{\Delta}\cdot\hat{\bm{X}}}$, which appears in $\Dop_+(E)$ and $\Dop_5(E)$, and the Gaussian of Eq.~\eqref{eq:Gauss},
\begin{align}
    &e^{i\bm{\Delta}\cdot\bm{r}} = \sum_{L\nu}i^L(2L + 1)C^L_\nu(\bm{e}_\Delta)^*j_L(|\bm{\Delta}|r)C^L_\nu(\bm{e}_r)\,,
    \label{eq:rayleigh} \\
    &\frac{1}{(\gamma\Mcl\sqrt\pi)^3}\exp[- \frac{|\bm{p} - \pket(E)|^2}{(\gamma\Mcl)^2}] = 
    \label{eq:Gaussexp} \\
    &\qquad \sum_{\lambda, \mu}(2\lambda + 1)w_\lambda\left(p, |\pket(E)|\right)C^\lambda_\mu(\bm{e}_{\mathrm{ket}})^*C^\lambda_\mu(\bm{e}_p)\,, \nonumber
\end{align}
with $|\nu| \le L$ and $|\mu| \le \lambda$.  The Racah-normalized spherical harmonics are $C^\lambda_\mu = \sqrt{4\pi/(2\lambda + 1)}Y_{\lambda\mu}$, where $Y_{\lambda\mu}$ are the spherical harmonics in the Condon--Shortley convention.  The unit vectors $\bm{e}_r$, $\bm{e}_p$, $\bm{e}_\Delta$, and $\bm{e}_{\mathrm{ket}}$ are the directions of $\bm{r}$, $\bm{p}$, $\bm{\Delta}$, and $\pket(E)$, and the symbols $r$ and $p$ denote $|\bm{r}|$ and $|\bm{p}|$.  The radial weight is
\begin{equation}
\begin{gathered}
    w_\lambda(p, p_0) = \frac{e^{- (p - p_0)^2 / (\gamma\Mcl)^2}}{(\gamma\Mcl\sqrt\pi)^3}e^{-\zeta}i_\lambda(\zeta)\,, \\
    \zeta = \frac{2pp_0}{(\gamma\Mcl)^2}\,,
\end{gathered}
\label{eq:wlambda}
\end{equation}
where $i_\lambda$ is the modified spherical Bessel function of the first kind.  For the GTMDs, all values of $\mu$ contribute, because $\pket(E)$ has the transverse part $\kp - \Dp/2$.  The angular factors follow from the algebra of the grand spin~\cite{Varshalovich:1988ifq}.  The grand spin and the parity restrict the multipoles to $|\lambda - L| \le 2$, and we truncate the sum over $L$ at $L_{\max}$.

Each sum over the occupied states in Eqs.~\eqref{eq:om0} and \eqref{eq:om1} equals minus the sum over the non-occupied states~\cite{Diakonov:1996sr, Pobylitsa:1998tk}.  As in Ref.~\cite{Minamiguchi:2026qcb}, we show the representation with the occupied states for $x \ge 0$ and the representation with the non-occupied states for $x < 0$.

\section{Model parameters}\label{app:settings}

The parameters, the box, the self-consistent profile, $\Mcl$, and $I$ are those of Ref.~\cite{Minamiguchi:2026qcb}.  The constituent quark mass is $M = 350$~MeV, and the pion mass is zero.  The Pauli--Villars masses are fixed by $f_\pi = 93$~MeV and by the chiral condensate $- (286.5~\mathrm{MeV})^3$ of one quark flavor.  The box has the radius $D = 30 / M$, and the basis contains the states with momenta up to $20M$.  The soliton has $\Mcl = 1061.8$~MeV and $I = 7.694\times10^{-3}$~MeV$^{-1}$.  The condition $\gamma\Mcl \ge \pi / D$ gives $\gamma \ge 0.0345$, and we use $\gamma = 0.04$.

For the GTMDs of Sec.~\ref{sec:results-gtmd}, we truncate the multipole sums at $L_{\max} = 32$ and convolve some sea contributions in $x$ (Appendix~\ref{app:numerics}).  For the Wigner distributions, we compute $F^\upd_{1, 1}$, $F^\upmd_{1, 4}$, and $G^\upd_{1, 1}$ at $\xi = 0$ and $x = \pm 0.1$ (Appendix~\ref{app:wigner}).

For the integrated OAM and SOC, the integrals over $\bm b_\perp$ in Eqs.~\eqref{eq:Lq} and \eqref{eq:Cz} give~\cite{Lorce:2011kd, Lorce:2014mxa}
\begin{align}
    L^\upmd &= -\int dx\int d^2\kp\,\frac{|\kp|^2}{M_N^2}F^\upmd_{1, 4}(x, 0, |\kp|^2, 0, 0)\,,
\label{eq:Lkint}\\
    C^\upd_z &= \int dx\int d^2\kp\,\frac{|\kp|^2}{M_N^2}G^\upd_{1, 1}(x, 0, |\kp|^2, 0, 0)\,.
\label{eq:Czkint}
\end{align}
Because $x$ enters $\Dop_+(E)$ and $\Dop_5(E)$ only through the third component of $\pket(E)$, the integral over $x$ of the Gaussian of Eq.~\eqref{eq:Gauss} gives
\begin{equation}
\begin{aligned}
    \int dx\,\Dop_+(E) &= \frac{\Gamma_+}{\pi\gamma^2\Mcl^3}e^{i\bm{\Delta}\cdot\hat{\bm{X}}} \\
    &\qquad\times\exp[-\frac{|\hat{\bm{p}}_\perp - \kp + \Dp/2|^2}{(\gamma\Mcl)^2}]\,,
\end{aligned}
\label{eq:xint}
\end{equation}
and the same expression with $\Gamma_5$ for $\Dop_5(E)$.  The right-hand side does not depend on $E$, and the derivative term in Eq.~\eqref{eq:Sdef} vanishes after the integral over $x$.  We therefore perform the integrals over $x$ in Eqs.~\eqref{eq:Lkint} and \eqref{eq:Czkint} before the sums over the quark states.  For the integrals over $\kp$, we use 80 values of $|\kp|$ up to $10M$ and $L_{\max} = 8$.  Because $F_{1, 4}$ and $G_{1, 1}$, and the GPD $E^\umd$ for $J^\umd$, enter the correlators with a factor proportional to $\Dp$, they cannot be extracted at $\Dp = 0$.  We therefore compute them at $\xi = 0$ and $-t = 10^{-4}$~GeV$^2$ instead of $t = 0$.

\section{Numerical checks}\label{app:checks}

We check the calculation in four ways.  First, in a small basis, we evaluate Eqs.~\eqref{eq:om0} and \eqref{eq:om1} in two ways: with the multipole expansions of Eqs.~\eqref{eq:rayleigh} and \eqref{eq:Gaussexp}, as in the production calculation, and with a numerical integration over the direction of the quark momentum without these expansions.  
The two evaluations of each coefficient agree within a relative difference of $3 \times 10^{-10}$.
Second, the integral of $F^\upd_{1, 1}$ over $\kp$ at $\xi = 0$ and $\Delta = 0$ gives the unpolarized GPD~\cite{Meissner:2009ww, Minamiguchi:2026qcb},
\begin{equation}
    \int d^2\kp\,F^\upd_{1, 1}(x, 0, |\kp|^2, 0, 0) = H^\upd(x, 0, 0)\,,
\label{eq:checkH}
\end{equation}
where both sides are smeared with the same $\gamma$.  For the total, the valence contribution, and the sea contribution, the two sides agree within $2\times10^{-5}$ of the maximum of $H^\upd(x, 0, 0)$.  Third, at $\xi = 0.20$ and $t = -0.30$~GeV$^2$, the sum rules
\begin{align}
    &\int dx\int d^2\kp\,F^\upd_{1,1} = F_1^\upd(t)\,,
\label{eq:sumF1}\\
    &2\int dx\int d^2\kp\left[\frac{\kp\cdot\Dp}{|\Dp|^2}F^\upd_{1,2}+F^\upd_{1,3}\right] = G_M^\upd(t)\,,
\label{eq:sumGM}
\end{align}
which hold at leading order in $1/N_c$~\cite{Meissner:2009ww}, are satisfied within $3\times10^{-5}$ relative.  Here $F_1^\upd$ and $G_M^\upd$ are the Dirac and magnetic form factors of the $u+d$ current.  Fourth, $L^\upd$ of Eq.~\eqref{eq:Lq} agrees within $3\times10^{-3}$ relative with the integral over $x$ of the distribution of the OAM of Ref.~\cite{Wakamatsu:1999nj}.  We compute this distribution with the method of Ref.~\cite{Minamiguchi:2026uaq}.
\section{Numerical evaluation of the Wigner distributions}\label{app:wigner}

We evaluate Eq.~\eqref{eq:wignerbessel} with ten values of $-t$ up to $1$~GeV$^2$, fourteen values of $|\kp|$ up to $1$~GeV, and eight values of $\phi$, four for $F^\upmd_{1, 4}$.  For $G^\upd_{1, 1}$ and $F^\upmd_{1, 4}$, we replace $-t = 0$ by $10^{-4}$~GeV$^2$.
Only the terms with $\kappa \le 1$ enter the figures.  At each value of $|\Dp|$, we keep every multipole $L$ for which $j_L(|\Dp|r)$ reaches half of its first maximum within $r \le D$~\cite{Minamiguchi:2026qcb}.  This criterion gives $L_{\max} = 86$ at $|\Dp| = 1$~GeV.
We interpolate each $a_\kappa$ with a cubic spline in $t$ up to $\Delta_{\max} = 1$~GeV and continue $a_\kappa$ beyond $\Delta_{\max}$ as
\begin{equation}
    a_\kappa = \left[a_\kappa(t_{\max}) + C_\kappa\,(t - t_{\max})\right]e^{\beta(t - t_{\max})}\,,
\label{eq:extrap}
\end{equation}
where $t_{\max} = -\Delta_{\max}^2$ and $\beta = 2$~GeV$^{-2}$.  The coefficient $C_\kappa = a_\kappa'(t_{\max}) - \beta\,a_\kappa(t_{\max})$ makes $a_\kappa$ and its derivative continuous.  
We integrate up to $|\Dp| = 3$~GeV.
Because $d^3L^\upd/(dx\,d|\bm b_\perp|\,d|\kp|)$ at $|\kp| > 0.75$~GeV is sensitive to this continuation, Fig.~\ref{fig:wignerUL_upd} shows only $|\kp| \le 0.75$~GeV.
The computed range, $-t \le 1$~GeV$^2$, is not necessarily sufficient for these Wigner distributions, and a more accurate evaluation without this continuation requires the GTMDs over a wider range of $t$.

\section{Convolution in $x$ and truncation of the multipole sums}\label{app:numerics}

\begin{table}
\caption{Values of $\gamma_x$.  The value $0.04$, which equals $\gamma$, means that the sea contribution is not convolved.  The sea contributions of the other GTMDs are not convolved.}
\label{tab:gammax}
\begin{ruledtabular}
\begin{tabular}{ccc}
GTMD & $\xi=0.20$ & $\xi=0$ \\
\hline
$F^\upd_{1,1}$ & $0.04$ & $0.05$ \\
$F^\umd_{1,1}$ & $0.06$ & $0.05$ \\
$F^\upd_{1,2}$ & $0.05$ & $0.05$ \\
$F^\upd_{1,4}$ & $0.06$ & $0.04$ \\
$G^\umd_{1,1}$ & $0.05$ & $0.04$ \\
$G^\upd_{1,3}$ & $0.07$ & $0.05$ \\
$G^\umd_{1,3}$ & $0.05$ & $0.04$ \\
$G^\upd_{1,4}$ & $0.06$ & $0.06$ \\
\end{tabular}
\end{ruledtabular}
\end{table}

The sea contributions of some GTMDs oscillate in $x$ on the scale of $\gamma$.
We convolve each of these sea contributions in $x$ with the normalized Gaussian $e^{-y^2/(\gamma_x^2 - \gamma^2)} / \sqrt{\pi(\gamma_x^2 - \gamma^2)}$ of the shift $y$.
The value of $\gamma_x$ in Table~\ref{tab:gammax} is the smallest value that removes the oscillation.
In Eq.~\eqref{eq:Gauss}, $x$ appears only in the third component $(x + \xi)\Mcl - E$ of $\pket(E)$ in Eq.~\eqref{eq:pket}.
A shift $y$ of $x$ thus moves the Gaussian of Eq.~\eqref{eq:Gauss} by $y\Mcl$ in the third component of the momentum.
The convolution in $x$ therefore changes the width of this Gaussian in the third component from $\gamma\Mcl$ to $\sqrt{\gamma^2 + (\gamma_x^2 - \gamma^2)}\,\Mcl = \gamma_x\Mcl$, while the width in the transverse components remains $\gamma\Mcl$.

Increasing $L_{\max}$ from $32$ to $47$ changes $F^\upd_{1, 1}$ at $\xi = 0.20$ by less than $2\times10^{-3}$ of its maximum.  The value $47$ is given by the criterion of Appendix~\ref{app:wigner} at $t = -0.30$~GeV$^2$.

\bibliography{paper_gtmd}

\begin{thebibliography}{69}%
\makeatletter
\providecommand \@ifxundefined [1]{%
 \@ifx{#1\undefined}
}%
\providecommand \@ifnum [1]{%
 \ifnum #1\expandafter \@firstoftwo
 \else \expandafter \@secondoftwo
 \fi
}%
\providecommand \@ifx [1]{%
 \ifx #1\expandafter \@firstoftwo
 \else \expandafter \@secondoftwo
 \fi
}%
\providecommand \natexlab [1]{#1}%
\providecommand \enquote  [1]{``#1''}%
\providecommand \bibnamefont  [1]{#1}%
\providecommand \bibfnamefont [1]{#1}%
\providecommand \citenamefont [1]{#1}%
\providecommand \href@noop [0]{\@secondoftwo}%
\providecommand \href [0]{\begingroup \@sanitize@url \@href}%
\providecommand \@href[1]{\@@startlink{#1}\@@href}%
\providecommand \@@href[1]{\endgroup#1\@@endlink}%
\providecommand \@sanitize@url [0]{\catcode `\\12\catcode `\$12\catcode `\&12\catcode `\#12\catcode `\^12\catcode `\_12\catcode `\%12\relax}%
\providecommand \@@startlink[1]{}%
\providecommand \@@endlink[0]{}%
\providecommand \url  [0]{\begingroup\@sanitize@url \@url }%
\providecommand \@url [1]{\endgroup\@href {#1}{\urlprefix }}%
\providecommand \urlprefix  [0]{URL }%
\providecommand \Eprint [0]{\href }%
\providecommand \doibase [0]{https://doi.org/}%
\providecommand \selectlanguage [0]{\@gobble}%
\providecommand \bibinfo  [0]{\@secondoftwo}%
\providecommand \bibfield  [0]{\@secondoftwo}%
\providecommand \translation [1]{[#1]}%
\providecommand \BibitemOpen [0]{}%
\providecommand \bibitemStop [0]{}%
\providecommand \bibitemNoStop [0]{.\EOS\space}%
\providecommand \EOS [0]{\spacefactor3000\relax}%
\providecommand \BibitemShut  [1]{\csname bibitem#1\endcsname}%
\let\auto@bib@innerbib\@empty
\bibitem [{\citenamefont {Meissner}\ \emph {et~al.}(2009)\citenamefont {Meissner}, \citenamefont {Metz},\ and\ \citenamefont {Schlegel}}]{Meissner:2009ww}%
  \BibitemOpen
  \bibfield  {author} {\bibinfo {author} {\bibfnamefont {S.}~\bibnamefont {Meissner}}, \bibinfo {author} {\bibfnamefont {A.}~\bibnamefont {Metz}},\ and\ \bibinfo {author} {\bibfnamefont {M.}~\bibnamefont {Schlegel}},\ }\href {https://doi.org/10.1088/1126-6708/2009/08/056} {\bibfield  {journal} {\bibinfo  {journal} {JHEP}\ }\textbf {\bibinfo {volume} {08}},\ \bibinfo {pages} {056}},\ \Eprint {https://arxiv.org/abs/0906.5323} {arXiv:0906.5323 [hep-ph]} \BibitemShut {NoStop}%
\bibitem [{\citenamefont {M{\"u}ller}\ \emph {et~al.}(1994)\citenamefont {M{\"u}ller}, \citenamefont {Robaschik}, \citenamefont {Geyer}, \citenamefont {Dittes},\ and\ \citenamefont {Ho{\v{r}}ej{\v{s}}i}}]{Muller:1994ses}%
  \BibitemOpen
  \bibfield  {author} {\bibinfo {author} {\bibfnamefont {D.}~\bibnamefont {M{\"u}ller}}, \bibinfo {author} {\bibfnamefont {D.}~\bibnamefont {Robaschik}}, \bibinfo {author} {\bibfnamefont {B.}~\bibnamefont {Geyer}}, \bibinfo {author} {\bibfnamefont {F.~M.}\ \bibnamefont {Dittes}},\ and\ \bibinfo {author} {\bibfnamefont {J.}~\bibnamefont {Ho{\v{r}}ej{\v{s}}i}},\ }\href {https://doi.org/10.1002/prop.2190420202} {\bibfield  {journal} {\bibinfo  {journal} {Fortsch. Phys.}\ }\textbf {\bibinfo {volume} {42}},\ \bibinfo {pages} {101} (\bibinfo {year} {1994})},\ \Eprint {https://arxiv.org/abs/hep-ph/9812448} {arXiv:hep-ph/9812448} \BibitemShut {NoStop}%
\bibitem [{\citenamefont {Ji}(1997)}]{Ji:1996ek}%
  \BibitemOpen
  \bibfield  {author} {\bibinfo {author} {\bibfnamefont {X.-D.}\ \bibnamefont {Ji}},\ }\href {https://doi.org/10.1103/PhysRevLett.78.610} {\bibfield  {journal} {\bibinfo  {journal} {Phys. Rev. Lett.}\ }\textbf {\bibinfo {volume} {78}},\ \bibinfo {pages} {610} (\bibinfo {year} {1997})},\ \Eprint {https://arxiv.org/abs/hep-ph/9603249} {arXiv:hep-ph/9603249} \BibitemShut {NoStop}%
\bibitem [{\citenamefont {Radyushkin}(1996)}]{Radyushkin:1996nd}%
  \BibitemOpen
  \bibfield  {author} {\bibinfo {author} {\bibfnamefont {A.~V.}\ \bibnamefont {Radyushkin}},\ }\href {https://doi.org/10.1016/0370-2693(96)00528-X} {\bibfield  {journal} {\bibinfo  {journal} {Phys. Lett. B}\ }\textbf {\bibinfo {volume} {380}},\ \bibinfo {pages} {417} (\bibinfo {year} {1996})},\ \Eprint {https://arxiv.org/abs/hep-ph/9604317} {arXiv:hep-ph/9604317} \BibitemShut {NoStop}%
\bibitem [{\citenamefont {Diehl}(2003)}]{Diehl:2003ny}%
  \BibitemOpen
  \bibfield  {author} {\bibinfo {author} {\bibfnamefont {M.}~\bibnamefont {Diehl}},\ }\href {https://doi.org/10.1016/j.physrep.2003.08.002} {\bibfield  {journal} {\bibinfo  {journal} {Phys. Rept.}\ }\textbf {\bibinfo {volume} {388}},\ \bibinfo {pages} {41} (\bibinfo {year} {2003})},\ \Eprint {https://arxiv.org/abs/hep-ph/0307382} {arXiv:hep-ph/0307382} \BibitemShut {NoStop}%
\bibitem [{\citenamefont {Mulders}\ and\ \citenamefont {Tangerman}(1996)}]{Mulders:1995dh}%
  \BibitemOpen
  \bibfield  {author} {\bibinfo {author} {\bibfnamefont {P.~J.}\ \bibnamefont {Mulders}}\ and\ \bibinfo {author} {\bibfnamefont {R.~D.}\ \bibnamefont {Tangerman}},\ }\href {https://doi.org/10.1016/0550-3213(95)00632-X} {\bibfield  {journal} {\bibinfo  {journal} {Nucl. Phys. B}\ }\textbf {\bibinfo {volume} {461}},\ \bibinfo {pages} {197} (\bibinfo {year} {1996})},\ \bibinfo {note} {[Erratum: Nucl.Phys.B 484, 538--540 (1997)]},\ \Eprint {https://arxiv.org/abs/hep-ph/9510301} {arXiv:hep-ph/9510301} \BibitemShut {NoStop}%
\bibitem [{\citenamefont {Bacchetta}\ \emph {et~al.}(2007)\citenamefont {Bacchetta}, \citenamefont {Diehl}, \citenamefont {Goeke}, \citenamefont {Metz}, \citenamefont {Mulders},\ and\ \citenamefont {Schlegel}}]{Bacchetta:2006tn}%
  \BibitemOpen
  \bibfield  {author} {\bibinfo {author} {\bibfnamefont {A.}~\bibnamefont {Bacchetta}}, \bibinfo {author} {\bibfnamefont {M.}~\bibnamefont {Diehl}}, \bibinfo {author} {\bibfnamefont {K.}~\bibnamefont {Goeke}}, \bibinfo {author} {\bibfnamefont {A.}~\bibnamefont {Metz}}, \bibinfo {author} {\bibfnamefont {P.~J.}\ \bibnamefont {Mulders}},\ and\ \bibinfo {author} {\bibfnamefont {M.}~\bibnamefont {Schlegel}},\ }\href {https://doi.org/10.1088/1126-6708/2007/02/093} {\bibfield  {journal} {\bibinfo  {journal} {JHEP}\ }\textbf {\bibinfo {volume} {02}},\ \bibinfo {pages} {093}},\ \Eprint {https://arxiv.org/abs/hep-ph/0611265} {arXiv:hep-ph/0611265} \BibitemShut {NoStop}%
\bibitem [{\citenamefont {Boussarie}\ \emph {et~al.}(2023)\citenamefont {Boussarie} \emph {et~al.}}]{Boussarie:2023izj}%
  \BibitemOpen
  \bibfield  {author} {\bibinfo {author} {\bibfnamefont {R.}~\bibnamefont {Boussarie}} \emph {et~al.},\ }\href@noop {} {\  (\bibinfo {year} {2023})},\ \Eprint {https://arxiv.org/abs/2304.03302} {arXiv:2304.03302 [hep-ph]} \BibitemShut {NoStop}%
\bibitem [{\citenamefont {Ji}(2003)}]{Ji:2003ak}%
  \BibitemOpen
  \bibfield  {author} {\bibinfo {author} {\bibfnamefont {X.-d.}\ \bibnamefont {Ji}},\ }\href {https://doi.org/10.1103/PhysRevLett.91.062001} {\bibfield  {journal} {\bibinfo  {journal} {Phys. Rev. Lett.}\ }\textbf {\bibinfo {volume} {91}},\ \bibinfo {pages} {062001} (\bibinfo {year} {2003})},\ \Eprint {https://arxiv.org/abs/hep-ph/0304037} {arXiv:hep-ph/0304037} \BibitemShut {NoStop}%
\bibitem [{\citenamefont {Belitsky}\ \emph {et~al.}(2004)\citenamefont {Belitsky}, \citenamefont {Ji},\ and\ \citenamefont {Yuan}}]{Belitsky:2003nz}%
  \BibitemOpen
  \bibfield  {author} {\bibinfo {author} {\bibfnamefont {A.~V.}\ \bibnamefont {Belitsky}}, \bibinfo {author} {\bibfnamefont {X.-d.}\ \bibnamefont {Ji}},\ and\ \bibinfo {author} {\bibfnamefont {F.}~\bibnamefont {Yuan}},\ }\href {https://doi.org/10.1103/PhysRevD.69.074014} {\bibfield  {journal} {\bibinfo  {journal} {Phys. Rev. D}\ }\textbf {\bibinfo {volume} {69}},\ \bibinfo {pages} {074014} (\bibinfo {year} {2004})},\ \Eprint {https://arxiv.org/abs/hep-ph/0307383} {arXiv:hep-ph/0307383} \BibitemShut {NoStop}%
\bibitem [{\citenamefont {Lorce}\ and\ \citenamefont {Pasquini}(2011)}]{Lorce:2011kd}%
  \BibitemOpen
  \bibfield  {author} {\bibinfo {author} {\bibfnamefont {C.}~\bibnamefont {Lorce}}\ and\ \bibinfo {author} {\bibfnamefont {B.}~\bibnamefont {Pasquini}},\ }\href {https://doi.org/10.1103/PhysRevD.84.014015} {\bibfield  {journal} {\bibinfo  {journal} {Phys. Rev. D}\ }\textbf {\bibinfo {volume} {84}},\ \bibinfo {pages} {014015} (\bibinfo {year} {2011})},\ \Eprint {https://arxiv.org/abs/1106.0139} {arXiv:1106.0139 [hep-ph]} \BibitemShut {NoStop}%
\bibitem [{\citenamefont {Hatta}(2012)}]{Hatta:2011ku}%
  \BibitemOpen
  \bibfield  {author} {\bibinfo {author} {\bibfnamefont {Y.}~\bibnamefont {Hatta}},\ }\href {https://doi.org/10.1016/j.physletb.2012.01.024} {\bibfield  {journal} {\bibinfo  {journal} {Phys. Lett. B}\ }\textbf {\bibinfo {volume} {708}},\ \bibinfo {pages} {186} (\bibinfo {year} {2012})},\ \Eprint {https://arxiv.org/abs/1111.3547} {arXiv:1111.3547 [hep-ph]} \BibitemShut {NoStop}%
\bibitem [{\citenamefont {Lorc{\'e}}(2014)}]{Lorce:2014mxa}%
  \BibitemOpen
  \bibfield  {author} {\bibinfo {author} {\bibfnamefont {C.}~\bibnamefont {Lorc{\'e}}},\ }\href {https://doi.org/10.1016/j.physletb.2014.06.068} {\bibfield  {journal} {\bibinfo  {journal} {Phys. Lett. B}\ }\textbf {\bibinfo {volume} {735}},\ \bibinfo {pages} {344} (\bibinfo {year} {2014})},\ \Eprint {https://arxiv.org/abs/1401.7784} {arXiv:1401.7784 [hep-ph]} \BibitemShut {NoStop}%
\bibitem [{\citenamefont {Hatta}\ \emph {et~al.}(2016)\citenamefont {Hatta}, \citenamefont {Xiao},\ and\ \citenamefont {Yuan}}]{Hatta:2016dxp}%
  \BibitemOpen
  \bibfield  {author} {\bibinfo {author} {\bibfnamefont {Y.}~\bibnamefont {Hatta}}, \bibinfo {author} {\bibfnamefont {B.-W.}\ \bibnamefont {Xiao}},\ and\ \bibinfo {author} {\bibfnamefont {F.}~\bibnamefont {Yuan}},\ }\href {https://doi.org/10.1103/PhysRevLett.116.202301} {\bibfield  {journal} {\bibinfo  {journal} {Phys. Rev. Lett.}\ }\textbf {\bibinfo {volume} {116}},\ \bibinfo {pages} {202301} (\bibinfo {year} {2016})},\ \Eprint {https://arxiv.org/abs/1601.01585} {arXiv:1601.01585 [hep-ph]} \BibitemShut {NoStop}%
\bibitem [{\citenamefont {Bhattacharya}\ \emph {et~al.}(2017)\citenamefont {Bhattacharya}, \citenamefont {Metz},\ and\ \citenamefont {Zhou}}]{Bhattacharya:2017bvs}%
  \BibitemOpen
  \bibfield  {author} {\bibinfo {author} {\bibfnamefont {S.}~\bibnamefont {Bhattacharya}}, \bibinfo {author} {\bibfnamefont {A.}~\bibnamefont {Metz}},\ and\ \bibinfo {author} {\bibfnamefont {J.}~\bibnamefont {Zhou}},\ }\href {https://doi.org/10.1016/j.physletb.2017.05.081} {\bibfield  {journal} {\bibinfo  {journal} {Phys. Lett. B}\ }\textbf {\bibinfo {volume} {771}},\ \bibinfo {pages} {396} (\bibinfo {year} {2017})},\ \bibinfo {note} {[Erratum: Phys.Lett.B 810, 135866 (2020)]},\ \Eprint {https://arxiv.org/abs/1702.04387} {arXiv:1702.04387 [hep-ph]} \BibitemShut {NoStop}%
\bibitem [{\citenamefont {Bhattacharya}\ \emph {et~al.}(2024{\natexlab{a}})\citenamefont {Bhattacharya}, \citenamefont {Zheng},\ and\ \citenamefont {Zhou}}]{Bhattacharya:2023hbq}%
  \BibitemOpen
  \bibfield  {author} {\bibinfo {author} {\bibfnamefont {S.}~\bibnamefont {Bhattacharya}}, \bibinfo {author} {\bibfnamefont {D.}~\bibnamefont {Zheng}},\ and\ \bibinfo {author} {\bibfnamefont {J.}~\bibnamefont {Zhou}},\ }\href {https://doi.org/10.1103/PhysRevLett.133.051901} {\bibfield  {journal} {\bibinfo  {journal} {Phys. Rev. Lett.}\ }\textbf {\bibinfo {volume} {133}},\ \bibinfo {pages} {051901} (\bibinfo {year} {2024}{\natexlab{a}})},\ \Eprint {https://arxiv.org/abs/2312.01309} {arXiv:2312.01309 [hep-ph]} \BibitemShut {NoStop}%
\bibitem [{\citenamefont {Bhattacharya}\ \emph {et~al.}(2026)\citenamefont {Bhattacharya}, \citenamefont {DeAngelo}, \citenamefont {Yang}, \citenamefont {Zheng},\ and\ \citenamefont {Zhou}}]{Bhattacharya:2026qnd}%
  \BibitemOpen
  \bibfield  {author} {\bibinfo {author} {\bibfnamefont {S.}~\bibnamefont {Bhattacharya}}, \bibinfo {author} {\bibfnamefont {D.}~\bibnamefont {DeAngelo}}, \bibinfo {author} {\bibfnamefont {L.}~\bibnamefont {Yang}}, \bibinfo {author} {\bibfnamefont {D.}~\bibnamefont {Zheng}},\ and\ \bibinfo {author} {\bibfnamefont {J.}~\bibnamefont {Zhou}},\ }\href {https://doi.org/10.1103/2qjg-mlxl} {\bibfield  {journal} {\bibinfo  {journal} {Phys. Rev. D}\ }\textbf {\bibinfo {volume} {114}},\ \bibinfo {pages} {054002} (\bibinfo {year} {2026})},\ \Eprint {https://arxiv.org/abs/2601.17506} {arXiv:2601.17506 [hep-ph]} \BibitemShut {NoStop}%
\bibitem [{\citenamefont {Engelhardt}(2017)}]{Engelhardt:2017miy}%
  \BibitemOpen
  \bibfield  {author} {\bibinfo {author} {\bibfnamefont {M.}~\bibnamefont {Engelhardt}},\ }\href {https://doi.org/10.1103/PhysRevD.95.094505} {\bibfield  {journal} {\bibinfo  {journal} {Phys. Rev. D}\ }\textbf {\bibinfo {volume} {95}},\ \bibinfo {pages} {094505} (\bibinfo {year} {2017})},\ \Eprint {https://arxiv.org/abs/1701.01536} {arXiv:1701.01536 [hep-lat]} \BibitemShut {NoStop}%
\bibitem [{\citenamefont {Engelhardt}\ \emph {et~al.}(2020)\citenamefont {Engelhardt}, \citenamefont {Green}, \citenamefont {Hasan}, \citenamefont {Krieg}, \citenamefont {Meinel}, \citenamefont {Negele}, \citenamefont {Pochinsky},\ and\ \citenamefont {Syritsyn}}]{Engelhardt:2020qtg}%
  \BibitemOpen
  \bibfield  {author} {\bibinfo {author} {\bibfnamefont {M.}~\bibnamefont {Engelhardt}}, \bibinfo {author} {\bibfnamefont {J.~R.}\ \bibnamefont {Green}}, \bibinfo {author} {\bibfnamefont {N.}~\bibnamefont {Hasan}}, \bibinfo {author} {\bibfnamefont {S.}~\bibnamefont {Krieg}}, \bibinfo {author} {\bibfnamefont {S.}~\bibnamefont {Meinel}}, \bibinfo {author} {\bibfnamefont {J.}~\bibnamefont {Negele}}, \bibinfo {author} {\bibfnamefont {A.}~\bibnamefont {Pochinsky}},\ and\ \bibinfo {author} {\bibfnamefont {S.}~\bibnamefont {Syritsyn}},\ }\href {https://doi.org/10.1103/PhysRevD.102.074505} {\bibfield  {journal} {\bibinfo  {journal} {Phys. Rev. D}\ }\textbf {\bibinfo {volume} {102}},\ \bibinfo {pages} {074505} (\bibinfo {year} {2020})},\ \Eprint {https://arxiv.org/abs/2008.03660} {arXiv:2008.03660 [hep-lat]} \BibitemShut {NoStop}%
\bibitem [{\citenamefont {Bertone}\ \emph {et~al.}(2025)\citenamefont {Bertone}, \citenamefont {Echevarria}, \citenamefont {del Rio},\ and\ \citenamefont {Rodini}}]{Bertone:2025vgy}%
  \BibitemOpen
  \bibfield  {author} {\bibinfo {author} {\bibfnamefont {V.}~\bibnamefont {Bertone}}, \bibinfo {author} {\bibfnamefont {M.~G.}\ \bibnamefont {Echevarria}}, \bibinfo {author} {\bibfnamefont {{\'O}.}~\bibnamefont {del Rio}},\ and\ \bibinfo {author} {\bibfnamefont {S.}~\bibnamefont {Rodini}},\ }\href {https://doi.org/10.1007/JHEP05(2025)183} {\bibfield  {journal} {\bibinfo  {journal} {JHEP}\ }\textbf {\bibinfo {volume} {05}},\ \bibinfo {pages} {183}},\ \Eprint {https://arxiv.org/abs/2502.07576} {arXiv:2502.07576 [hep-ph]} \BibitemShut {NoStop}%
\bibitem [{\citenamefont {Beni{\'c}}\ \emph {et~al.}(2026)\citenamefont {Beni{\'c}}, \citenamefont {Hagiwara}, \citenamefont {{\v{S}}ari{\'c}},\ and\ \citenamefont {Vivoda}}]{Benic:2026idy}%
  \BibitemOpen
  \bibfield  {author} {\bibinfo {author} {\bibfnamefont {S.}~\bibnamefont {Beni{\'c}}}, \bibinfo {author} {\bibfnamefont {Y.}~\bibnamefont {Hagiwara}}, \bibinfo {author} {\bibfnamefont {B.}~\bibnamefont {{\v{S}}ari{\'c}}},\ and\ \bibinfo {author} {\bibfnamefont {E.~A.}\ \bibnamefont {Vivoda}},\ }\href {https://doi.org/10.1007/JHEP09(2026)142} {\bibfield  {journal} {\bibinfo  {journal} {JHEP}\ }\textbf {\bibinfo {volume} {09}},\ \bibinfo {pages} {142}},\ \Eprint {https://arxiv.org/abs/2603.06092} {arXiv:2603.06092 [hep-ph]} \BibitemShut {NoStop}%
\bibitem [{\citenamefont {Hatta}\ \emph {et~al.}(2017)\citenamefont {Hatta}, \citenamefont {Nakagawa}, \citenamefont {Yuan}, \citenamefont {Zhao},\ and\ \citenamefont {Xiao}}]{Hatta:2016aoc}%
  \BibitemOpen
  \bibfield  {author} {\bibinfo {author} {\bibfnamefont {Y.}~\bibnamefont {Hatta}}, \bibinfo {author} {\bibfnamefont {Y.}~\bibnamefont {Nakagawa}}, \bibinfo {author} {\bibfnamefont {F.}~\bibnamefont {Yuan}}, \bibinfo {author} {\bibfnamefont {Y.}~\bibnamefont {Zhao}},\ and\ \bibinfo {author} {\bibfnamefont {B.}~\bibnamefont {Xiao}},\ }\href {https://doi.org/10.1103/PhysRevD.95.114032} {\bibfield  {journal} {\bibinfo  {journal} {Phys. Rev. D}\ }\textbf {\bibinfo {volume} {95}},\ \bibinfo {pages} {114032} (\bibinfo {year} {2017})},\ \Eprint {https://arxiv.org/abs/1612.02445} {arXiv:1612.02445 [hep-ph]} \BibitemShut {NoStop}%
\bibitem [{\citenamefont {Boussarie}\ \emph {et~al.}(2019)\citenamefont {Boussarie}, \citenamefont {Hatta},\ and\ \citenamefont {Yuan}}]{Boussarie:2019icw}%
  \BibitemOpen
  \bibfield  {author} {\bibinfo {author} {\bibfnamefont {R.}~\bibnamefont {Boussarie}}, \bibinfo {author} {\bibfnamefont {Y.}~\bibnamefont {Hatta}},\ and\ \bibinfo {author} {\bibfnamefont {F.}~\bibnamefont {Yuan}},\ }\href {https://doi.org/10.1016/j.physletb.2019.134817} {\bibfield  {journal} {\bibinfo  {journal} {Phys. Lett. B}\ }\textbf {\bibinfo {volume} {797}},\ \bibinfo {pages} {134817} (\bibinfo {year} {2019})},\ \Eprint {https://arxiv.org/abs/1904.02693} {arXiv:1904.02693 [hep-ph]} \BibitemShut {NoStop}%
\bibitem [{\citenamefont {Kovchegov}\ and\ \citenamefont {Manley}(2024)}]{Kovchegov:2023yzd}%
  \BibitemOpen
  \bibfield  {author} {\bibinfo {author} {\bibfnamefont {Y.~V.}\ \bibnamefont {Kovchegov}}\ and\ \bibinfo {author} {\bibfnamefont {B.}~\bibnamefont {Manley}},\ }\href {https://doi.org/10.1007/JHEP02(2024)060} {\bibfield  {journal} {\bibinfo  {journal} {JHEP}\ }\textbf {\bibinfo {volume} {02}},\ \bibinfo {pages} {060}},\ \bibinfo {note} {[Erratum: JHEP 08, 140 (2024)]},\ \Eprint {https://arxiv.org/abs/2310.18404} {arXiv:2310.18404 [hep-ph]} \BibitemShut {NoStop}%
\bibitem [{\citenamefont {Manley}(2024)}]{Manley:2024pcl}%
  \BibitemOpen
  \bibfield  {author} {\bibinfo {author} {\bibfnamefont {B.}~\bibnamefont {Manley}},\ }\href {https://doi.org/10.1007/JHEP04(2024)055} {\bibfield  {journal} {\bibinfo  {journal} {JHEP}\ }\textbf {\bibinfo {volume} {04}},\ \bibinfo {pages} {055}},\ \bibinfo {note} {[Erratum: JHEP 09, 017 (2024)]},\ \Eprint {https://arxiv.org/abs/2401.05508} {arXiv:2401.05508 [hep-ph]} \BibitemShut {NoStop}%
\bibitem [{\citenamefont {Kovchegov}\ and\ \citenamefont {Manley}(2026)}]{Kovchegov:2026lik}%
  \BibitemOpen
  \bibfield  {author} {\bibinfo {author} {\bibfnamefont {Y.~V.}\ \bibnamefont {Kovchegov}}\ and\ \bibinfo {author} {\bibfnamefont {B.}~\bibnamefont {Manley}},\ }\href@noop {} {\  (\bibinfo {year} {2026})},\ \Eprint {https://arxiv.org/abs/2610.33255} {arXiv:2610.33255 [hep-ph]} \BibitemShut {NoStop}%
\bibitem [{\citenamefont {Bhattacharya}\ \emph {et~al.}(2024{\natexlab{b}})\citenamefont {Bhattacharya}, \citenamefont {Boussarie},\ and\ \citenamefont {Hatta}}]{Bhattacharya:2024sno}%
  \BibitemOpen
  \bibfield  {author} {\bibinfo {author} {\bibfnamefont {S.}~\bibnamefont {Bhattacharya}}, \bibinfo {author} {\bibfnamefont {R.}~\bibnamefont {Boussarie}},\ and\ \bibinfo {author} {\bibfnamefont {Y.}~\bibnamefont {Hatta}},\ }\href {https://doi.org/10.1016/j.physletb.2024.139134} {\bibfield  {journal} {\bibinfo  {journal} {Phys. Lett. B}\ }\textbf {\bibinfo {volume} {859}},\ \bibinfo {pages} {139134} (\bibinfo {year} {2024}{\natexlab{b}})},\ \Eprint {https://arxiv.org/abs/2404.04208} {arXiv:2404.04208 [hep-ph]} \BibitemShut {NoStop}%
\bibitem [{\citenamefont {Beni{\'c}}\ and\ \citenamefont {Hatta}(2025)}]{Benic:2025xpg}%
  \BibitemOpen
  \bibfield  {author} {\bibinfo {author} {\bibfnamefont {S.}~\bibnamefont {Beni{\'c}}}\ and\ \bibinfo {author} {\bibfnamefont {Y.}~\bibnamefont {Hatta}},\ }\href {https://doi.org/10.1016/j.physletb.2025.139709} {\bibfield  {journal} {\bibinfo  {journal} {Phys. Lett. B}\ }\textbf {\bibinfo {volume} {868}},\ \bibinfo {pages} {139709} (\bibinfo {year} {2025})},\ \Eprint {https://arxiv.org/abs/2505.05172} {arXiv:2505.05172 [hep-ph]} \BibitemShut {NoStop}%
\bibitem [{\citenamefont {Lorce}\ \emph {et~al.}(2011)\citenamefont {Lorce}, \citenamefont {Pasquini},\ and\ \citenamefont {Vanderhaeghen}}]{Lorce:2011dv}%
  \BibitemOpen
  \bibfield  {author} {\bibinfo {author} {\bibfnamefont {C.}~\bibnamefont {Lorce}}, \bibinfo {author} {\bibfnamefont {B.}~\bibnamefont {Pasquini}},\ and\ \bibinfo {author} {\bibfnamefont {M.}~\bibnamefont {Vanderhaeghen}},\ }\href {https://doi.org/10.1007/JHEP05(2011)041} {\bibfield  {journal} {\bibinfo  {journal} {JHEP}\ }\textbf {\bibinfo {volume} {05}},\ \bibinfo {pages} {041}},\ \Eprint {https://arxiv.org/abs/1102.4704} {arXiv:1102.4704 [hep-ph]} \BibitemShut {NoStop}%
\bibitem [{\citenamefont {Luan}\ and\ \citenamefont {Lu}(2024)}]{Luan:2024nwc}%
  \BibitemOpen
  \bibfield  {author} {\bibinfo {author} {\bibfnamefont {X.}~\bibnamefont {Luan}}\ and\ \bibinfo {author} {\bibfnamefont {Z.}~\bibnamefont {Lu}},\ }\href {https://doi.org/10.1103/PhysRevD.109.094016} {\bibfield  {journal} {\bibinfo  {journal} {Phys. Rev. D}\ }\textbf {\bibinfo {volume} {109}},\ \bibinfo {pages} {094016} (\bibinfo {year} {2024})},\ \Eprint {https://arxiv.org/abs/2401.15596} {arXiv:2401.15596 [hep-ph]} \BibitemShut {NoStop}%
\bibitem [{\citenamefont {Maynard}\ and\ \citenamefont {Schweitzer}(2026)}]{Maynard:2026atm}%
  \BibitemOpen
  \bibfield  {author} {\bibinfo {author} {\bibfnamefont {B.}~\bibnamefont {Maynard}}\ and\ \bibinfo {author} {\bibfnamefont {P.}~\bibnamefont {Schweitzer}},\ }\href {https://doi.org/10.1103/b6kc-fg77} {\bibfield  {journal} {\bibinfo  {journal} {Phys. Rev. D}\ }\textbf {\bibinfo {volume} {114}},\ \bibinfo {pages} {076007} (\bibinfo {year} {2026})},\ \Eprint {https://arxiv.org/abs/2605.06412} {arXiv:2605.06412 [hep-ph]} \BibitemShut {NoStop}%
\bibitem [{\citenamefont {Schweitzer}\ \emph {et~al.}(2013)\citenamefont {Schweitzer}, \citenamefont {Strikman},\ and\ \citenamefont {Weiss}}]{Schweitzer:2012hh}%
  \BibitemOpen
  \bibfield  {author} {\bibinfo {author} {\bibfnamefont {P.}~\bibnamefont {Schweitzer}}, \bibinfo {author} {\bibfnamefont {M.}~\bibnamefont {Strikman}},\ and\ \bibinfo {author} {\bibfnamefont {C.}~\bibnamefont {Weiss}},\ }\href {https://doi.org/10.1007/JHEP01(2013)163} {\bibfield  {journal} {\bibinfo  {journal} {JHEP}\ }\textbf {\bibinfo {volume} {01}},\ \bibinfo {pages} {163}},\ \Eprint {https://arxiv.org/abs/1210.1267} {arXiv:1210.1267 [hep-ph]} \BibitemShut {NoStop}%
\bibitem [{\citenamefont {Diakonov}\ \emph {et~al.}(1988)\citenamefont {Diakonov}, \citenamefont {Petrov},\ and\ \citenamefont {Pobylitsa}}]{Diakonov:1987ty}%
  \BibitemOpen
  \bibfield  {author} {\bibinfo {author} {\bibfnamefont {D.}~\bibnamefont {Diakonov}}, \bibinfo {author} {\bibfnamefont {V.~Y.}\ \bibnamefont {Petrov}},\ and\ \bibinfo {author} {\bibfnamefont {P.~V.}\ \bibnamefont {Pobylitsa}},\ }\href {https://doi.org/10.1016/0550-3213(88)90443-9} {\bibfield  {journal} {\bibinfo  {journal} {Nucl. Phys. B}\ }\textbf {\bibinfo {volume} {306}},\ \bibinfo {pages} {809} (\bibinfo {year} {1988})}\BibitemShut {NoStop}%
\bibitem [{\citenamefont {Reinhardt}\ and\ \citenamefont {Wunsch}(1988)}]{Reinhardt:1988fz}%
  \BibitemOpen
  \bibfield  {author} {\bibinfo {author} {\bibfnamefont {H.}~\bibnamefont {Reinhardt}}\ and\ \bibinfo {author} {\bibfnamefont {R.}~\bibnamefont {Wunsch}},\ }\href {https://doi.org/10.1016/0370-2693(88)91363-9} {\bibfield  {journal} {\bibinfo  {journal} {Phys. Lett. B}\ }\textbf {\bibinfo {volume} {215}},\ \bibinfo {pages} {577} (\bibinfo {year} {1988})}\BibitemShut {NoStop}%
\bibitem [{\citenamefont {Alkofer}\ \emph {et~al.}(1996)\citenamefont {Alkofer}, \citenamefont {Reinhardt},\ and\ \citenamefont {Weigel}}]{Alkofer:1994ph}%
  \BibitemOpen
  \bibfield  {author} {\bibinfo {author} {\bibfnamefont {R.}~\bibnamefont {Alkofer}}, \bibinfo {author} {\bibfnamefont {H.}~\bibnamefont {Reinhardt}},\ and\ \bibinfo {author} {\bibfnamefont {H.}~\bibnamefont {Weigel}},\ }\href {https://doi.org/10.1016/0370-1573(95)00018-6} {\bibfield  {journal} {\bibinfo  {journal} {Phys. Rept.}\ }\textbf {\bibinfo {volume} {265}},\ \bibinfo {pages} {139} (\bibinfo {year} {1996})},\ \Eprint {https://arxiv.org/abs/hep-ph/9501213} {arXiv:hep-ph/9501213} \BibitemShut {NoStop}%
\bibitem [{\citenamefont {Christov}\ \emph {et~al.}(1996)\citenamefont {Christov}, \citenamefont {Blotz}, \citenamefont {Kim}, \citenamefont {Pobylitsa}, \citenamefont {Watabe}, \citenamefont {Meissner}, \citenamefont {Ruiz~Arriola},\ and\ \citenamefont {Goeke}}]{Christov:1995vm}%
  \BibitemOpen
  \bibfield  {author} {\bibinfo {author} {\bibfnamefont {C.~V.}\ \bibnamefont {Christov}}, \bibinfo {author} {\bibfnamefont {A.}~\bibnamefont {Blotz}}, \bibinfo {author} {\bibfnamefont {H.-C.}\ \bibnamefont {Kim}}, \bibinfo {author} {\bibfnamefont {P.}~\bibnamefont {Pobylitsa}}, \bibinfo {author} {\bibfnamefont {T.}~\bibnamefont {Watabe}}, \bibinfo {author} {\bibfnamefont {T.}~\bibnamefont {Meissner}}, \bibinfo {author} {\bibfnamefont {E.}~\bibnamefont {Ruiz~Arriola}},\ and\ \bibinfo {author} {\bibfnamefont {K.}~\bibnamefont {Goeke}},\ }\href {https://doi.org/10.1016/0146-6410(96)00057-9} {\bibfield  {journal} {\bibinfo  {journal} {Prog. Part. Nucl. Phys.}\ }\textbf {\bibinfo {volume} {37}},\ \bibinfo {pages} {91} (\bibinfo {year} {1996})},\ \Eprint {https://arxiv.org/abs/hep-ph/9604441} {arXiv:hep-ph/9604441} \BibitemShut {NoStop}%
\bibitem [{\citenamefont {Diakonov}\ \emph {et~al.}(1997)\citenamefont {Diakonov}, \citenamefont {Petrov}, \citenamefont {Pobylitsa}, \citenamefont {Polyakov},\ and\ \citenamefont {Weiss}}]{Diakonov:1997vc}%
  \BibitemOpen
  \bibfield  {author} {\bibinfo {author} {\bibfnamefont {D.}~\bibnamefont {Diakonov}}, \bibinfo {author} {\bibfnamefont {V.~Y.}\ \bibnamefont {Petrov}}, \bibinfo {author} {\bibfnamefont {P.~V.}\ \bibnamefont {Pobylitsa}}, \bibinfo {author} {\bibfnamefont {M.~V.}\ \bibnamefont {Polyakov}},\ and\ \bibinfo {author} {\bibfnamefont {C.}~\bibnamefont {Weiss}},\ }\href {https://doi.org/10.1103/PhysRevD.56.4069} {\bibfield  {journal} {\bibinfo  {journal} {Phys. Rev. D}\ }\textbf {\bibinfo {volume} {56}},\ \bibinfo {pages} {4069} (\bibinfo {year} {1997})},\ \Eprint {https://arxiv.org/abs/hep-ph/9703420} {arXiv:hep-ph/9703420} \BibitemShut {NoStop}%
\bibitem [{\citenamefont {Weigel}\ \emph {et~al.}(1997{\natexlab{a}})\citenamefont {Weigel}, \citenamefont {Gamberg},\ and\ \citenamefont {Reinhardt}}]{Weigel:1996kw}%
  \BibitemOpen
  \bibfield  {author} {\bibinfo {author} {\bibfnamefont {H.}~\bibnamefont {Weigel}}, \bibinfo {author} {\bibfnamefont {L.~P.}\ \bibnamefont {Gamberg}},\ and\ \bibinfo {author} {\bibfnamefont {H.}~\bibnamefont {Reinhardt}},\ }\href {https://doi.org/10.1016/S0370-2693(97)00296-7} {\bibfield  {journal} {\bibinfo  {journal} {Phys. Lett. B}\ }\textbf {\bibinfo {volume} {399}},\ \bibinfo {pages} {287} (\bibinfo {year} {1997}{\natexlab{a}})},\ \Eprint {https://arxiv.org/abs/hep-ph/9604295} {arXiv:hep-ph/9604295} \BibitemShut {NoStop}%
\bibitem [{\citenamefont {Diakonov}\ \emph {et~al.}(1996)\citenamefont {Diakonov}, \citenamefont {Petrov}, \citenamefont {Pobylitsa}, \citenamefont {Polyakov},\ and\ \citenamefont {Weiss}}]{Diakonov:1996sr}%
  \BibitemOpen
  \bibfield  {author} {\bibinfo {author} {\bibfnamefont {D.}~\bibnamefont {Diakonov}}, \bibinfo {author} {\bibfnamefont {V.}~\bibnamefont {Petrov}}, \bibinfo {author} {\bibfnamefont {P.}~\bibnamefont {Pobylitsa}}, \bibinfo {author} {\bibfnamefont {M.~V.}\ \bibnamefont {Polyakov}},\ and\ \bibinfo {author} {\bibfnamefont {C.}~\bibnamefont {Weiss}},\ }\href {https://doi.org/10.1016/S0550-3213(96)00486-5} {\bibfield  {journal} {\bibinfo  {journal} {Nucl. Phys. B}\ }\textbf {\bibinfo {volume} {480}},\ \bibinfo {pages} {341} (\bibinfo {year} {1996})},\ \Eprint {https://arxiv.org/abs/hep-ph/9606314} {arXiv:hep-ph/9606314} \BibitemShut {NoStop}%
\bibitem [{\citenamefont {Weigel}\ \emph {et~al.}(1997{\natexlab{b}})\citenamefont {Weigel}, \citenamefont {Gamberg},\ and\ \citenamefont {Reinhardt}}]{Weigel:1996jh}%
  \BibitemOpen
  \bibfield  {author} {\bibinfo {author} {\bibfnamefont {H.}~\bibnamefont {Weigel}}, \bibinfo {author} {\bibfnamefont {L.~P.}\ \bibnamefont {Gamberg}},\ and\ \bibinfo {author} {\bibfnamefont {H.}~\bibnamefont {Reinhardt}},\ }\href {https://doi.org/10.1103/PhysRevD.55.6910} {\bibfield  {journal} {\bibinfo  {journal} {Phys. Rev. D}\ }\textbf {\bibinfo {volume} {55}},\ \bibinfo {pages} {6910} (\bibinfo {year} {1997}{\natexlab{b}})},\ \Eprint {https://arxiv.org/abs/hep-ph/9609226} {arXiv:hep-ph/9609226} \BibitemShut {NoStop}%
\bibitem [{\citenamefont {Weiss}\ and\ \citenamefont {Goeke}(1997)}]{Weiss:1997rt}%
  \BibitemOpen
  \bibfield  {author} {\bibinfo {author} {\bibfnamefont {C.}~\bibnamefont {Weiss}}\ and\ \bibinfo {author} {\bibfnamefont {K.}~\bibnamefont {Goeke}},\ }\href@noop {} {\  (\bibinfo {year} {1997})},\ \Eprint {https://arxiv.org/abs/hep-ph/9712447} {arXiv:hep-ph/9712447} \BibitemShut {NoStop}%
\bibitem [{\citenamefont {Wakamatsu}\ and\ \citenamefont {Kubota}(1998)}]{Wakamatsu:1997en}%
  \BibitemOpen
  \bibfield  {author} {\bibinfo {author} {\bibfnamefont {M.}~\bibnamefont {Wakamatsu}}\ and\ \bibinfo {author} {\bibfnamefont {T.}~\bibnamefont {Kubota}},\ }\href {https://doi.org/10.1103/PhysRevD.57.5755} {\bibfield  {journal} {\bibinfo  {journal} {Phys. Rev. D}\ }\textbf {\bibinfo {volume} {57}},\ \bibinfo {pages} {5755} (\bibinfo {year} {1998})},\ \Eprint {https://arxiv.org/abs/hep-ph/9707500} {arXiv:hep-ph/9707500} \BibitemShut {NoStop}%
\bibitem [{\citenamefont {Pobylitsa}\ \emph {et~al.}(1999)\citenamefont {Pobylitsa}, \citenamefont {Polyakov}, \citenamefont {Goeke}, \citenamefont {Watabe},\ and\ \citenamefont {Weiss}}]{Pobylitsa:1998tk}%
  \BibitemOpen
  \bibfield  {author} {\bibinfo {author} {\bibfnamefont {P.~V.}\ \bibnamefont {Pobylitsa}}, \bibinfo {author} {\bibfnamefont {M.~V.}\ \bibnamefont {Polyakov}}, \bibinfo {author} {\bibfnamefont {K.}~\bibnamefont {Goeke}}, \bibinfo {author} {\bibfnamefont {T.}~\bibnamefont {Watabe}},\ and\ \bibinfo {author} {\bibfnamefont {C.}~\bibnamefont {Weiss}},\ }\href {https://doi.org/10.1103/PhysRevD.59.034024} {\bibfield  {journal} {\bibinfo  {journal} {Phys. Rev. D}\ }\textbf {\bibinfo {volume} {59}},\ \bibinfo {pages} {034024} (\bibinfo {year} {1999})},\ \Eprint {https://arxiv.org/abs/hep-ph/9804436} {arXiv:hep-ph/9804436} \BibitemShut {NoStop}%
\bibitem [{\citenamefont {Wakamatsu}\ and\ \citenamefont {Kubota}(1999)}]{Wakamatsu:1998rx}%
  \BibitemOpen
  \bibfield  {author} {\bibinfo {author} {\bibfnamefont {M.}~\bibnamefont {Wakamatsu}}\ and\ \bibinfo {author} {\bibfnamefont {T.}~\bibnamefont {Kubota}},\ }\href {https://doi.org/10.1103/PhysRevD.60.034020} {\bibfield  {journal} {\bibinfo  {journal} {Phys. Rev. D}\ }\textbf {\bibinfo {volume} {60}},\ \bibinfo {pages} {034020} (\bibinfo {year} {1999})},\ \Eprint {https://arxiv.org/abs/hep-ph/9809443} {arXiv:hep-ph/9809443} \BibitemShut {NoStop}%
\bibitem [{\citenamefont {Weigel}\ \emph {et~al.}(1999)\citenamefont {Weigel}, \citenamefont {Ruiz~Arriola},\ and\ \citenamefont {Gamberg}}]{Weigel:1999pc}%
  \BibitemOpen
  \bibfield  {author} {\bibinfo {author} {\bibfnamefont {H.}~\bibnamefont {Weigel}}, \bibinfo {author} {\bibfnamefont {E.}~\bibnamefont {Ruiz~Arriola}},\ and\ \bibinfo {author} {\bibfnamefont {L.~P.}\ \bibnamefont {Gamberg}},\ }\href {https://doi.org/10.1016/S0550-3213(99)00426-5} {\bibfield  {journal} {\bibinfo  {journal} {Nucl. Phys. B}\ }\textbf {\bibinfo {volume} {560}},\ \bibinfo {pages} {383} (\bibinfo {year} {1999})},\ \Eprint {https://arxiv.org/abs/hep-ph/9905329} {arXiv:hep-ph/9905329} \BibitemShut {NoStop}%
\bibitem [{\citenamefont {Wakamatsu}\ and\ \citenamefont {Watabe}(2000)}]{Wakamatsu:1999nj}%
  \BibitemOpen
  \bibfield  {author} {\bibinfo {author} {\bibfnamefont {M.}~\bibnamefont {Wakamatsu}}\ and\ \bibinfo {author} {\bibfnamefont {T.}~\bibnamefont {Watabe}},\ }\href {https://doi.org/10.1103/PhysRevD.62.054009} {\bibfield  {journal} {\bibinfo  {journal} {Phys. Rev. D}\ }\textbf {\bibinfo {volume} {62}},\ \bibinfo {pages} {054009} (\bibinfo {year} {2000})},\ \Eprint {https://arxiv.org/abs/hep-ph/9912500} {arXiv:hep-ph/9912500} \BibitemShut {NoStop}%
\bibitem [{\citenamefont {Takyi}\ and\ \citenamefont {Weigel}(2019)}]{Takyi:2019ahv}%
  \BibitemOpen
  \bibfield  {author} {\bibinfo {author} {\bibfnamefont {I.}~\bibnamefont {Takyi}}\ and\ \bibinfo {author} {\bibfnamefont {H.}~\bibnamefont {Weigel}},\ }\href {https://doi.org/10.1140/epja/i2019-12806-3} {\bibfield  {journal} {\bibinfo  {journal} {Eur. Phys. J. A}\ }\textbf {\bibinfo {volume} {55}},\ \bibinfo {pages} {128} (\bibinfo {year} {2019})},\ \Eprint {https://arxiv.org/abs/1903.11435} {arXiv:1903.11435 [hep-ph]} \BibitemShut {NoStop}%
\bibitem [{\citenamefont {Petrov}\ \emph {et~al.}(1998)\citenamefont {Petrov}, \citenamefont {Pobylitsa}, \citenamefont {Polyakov}, \citenamefont {Bornig}, \citenamefont {Goeke},\ and\ \citenamefont {Weiss}}]{Petrov:1998kf}%
  \BibitemOpen
  \bibfield  {author} {\bibinfo {author} {\bibfnamefont {V.~Y.}\ \bibnamefont {Petrov}}, \bibinfo {author} {\bibfnamefont {P.~V.}\ \bibnamefont {Pobylitsa}}, \bibinfo {author} {\bibfnamefont {M.~V.}\ \bibnamefont {Polyakov}}, \bibinfo {author} {\bibfnamefont {I.}~\bibnamefont {Bornig}}, \bibinfo {author} {\bibfnamefont {K.}~\bibnamefont {Goeke}},\ and\ \bibinfo {author} {\bibfnamefont {C.}~\bibnamefont {Weiss}},\ }\href {https://doi.org/10.1103/PhysRevD.57.4325} {\bibfield  {journal} {\bibinfo  {journal} {Phys. Rev. D}\ }\textbf {\bibinfo {volume} {57}},\ \bibinfo {pages} {4325} (\bibinfo {year} {1998})},\ \Eprint {https://arxiv.org/abs/hep-ph/9710270} {arXiv:hep-ph/9710270} \BibitemShut {NoStop}%
\bibitem [{\citenamefont {Penttinen}\ \emph {et~al.}(2000)\citenamefont {Penttinen}, \citenamefont {Polyakov},\ and\ \citenamefont {Goeke}}]{Penttinen:1999th}%
  \BibitemOpen
  \bibfield  {author} {\bibinfo {author} {\bibfnamefont {M.}~\bibnamefont {Penttinen}}, \bibinfo {author} {\bibfnamefont {M.~V.}\ \bibnamefont {Polyakov}},\ and\ \bibinfo {author} {\bibfnamefont {K.}~\bibnamefont {Goeke}},\ }\href {https://doi.org/10.1103/PhysRevD.62.014024} {\bibfield  {journal} {\bibinfo  {journal} {Phys. Rev. D}\ }\textbf {\bibinfo {volume} {62}},\ \bibinfo {pages} {014024} (\bibinfo {year} {2000})},\ \Eprint {https://arxiv.org/abs/hep-ph/9909489} {arXiv:hep-ph/9909489} \BibitemShut {NoStop}%
\bibitem [{\citenamefont {Ossmann}\ \emph {et~al.}(2005)\citenamefont {Ossmann}, \citenamefont {Polyakov}, \citenamefont {Schweitzer}, \citenamefont {Urbano},\ and\ \citenamefont {Goeke}}]{Ossmann:2004bp}%
  \BibitemOpen
  \bibfield  {author} {\bibinfo {author} {\bibfnamefont {J.}~\bibnamefont {Ossmann}}, \bibinfo {author} {\bibfnamefont {M.~V.}\ \bibnamefont {Polyakov}}, \bibinfo {author} {\bibfnamefont {P.}~\bibnamefont {Schweitzer}}, \bibinfo {author} {\bibfnamefont {D.}~\bibnamefont {Urbano}},\ and\ \bibinfo {author} {\bibfnamefont {K.}~\bibnamefont {Goeke}},\ }\href {https://doi.org/10.1103/PhysRevD.71.034011} {\bibfield  {journal} {\bibinfo  {journal} {Phys. Rev. D}\ }\textbf {\bibinfo {volume} {71}},\ \bibinfo {pages} {034011} (\bibinfo {year} {2005})},\ \Eprint {https://arxiv.org/abs/hep-ph/0411172} {arXiv:hep-ph/0411172} \BibitemShut {NoStop}%
\bibitem [{\citenamefont {Wakamatsu}\ and\ \citenamefont {Tsujimoto}(2005)}]{Wakamatsu:2005vk}%
  \BibitemOpen
  \bibfield  {author} {\bibinfo {author} {\bibfnamefont {M.}~\bibnamefont {Wakamatsu}}\ and\ \bibinfo {author} {\bibfnamefont {H.}~\bibnamefont {Tsujimoto}},\ }\href {https://doi.org/10.1103/PhysRevD.71.074001} {\bibfield  {journal} {\bibinfo  {journal} {Phys. Rev. D}\ }\textbf {\bibinfo {volume} {71}},\ \bibinfo {pages} {074001} (\bibinfo {year} {2005})},\ \Eprint {https://arxiv.org/abs/hep-ph/0502030} {arXiv:hep-ph/0502030} \BibitemShut {NoStop}%
\bibitem [{\citenamefont {Pobylitsa}(2003)}]{Pobylitsa:2003ty}%
  \BibitemOpen
  \bibfield  {author} {\bibinfo {author} {\bibfnamefont {P.~V.}\ \bibnamefont {Pobylitsa}},\ }\href@noop {} {\  (\bibinfo {year} {2003})},\ \Eprint {https://arxiv.org/abs/hep-ph/0301236} {arXiv:hep-ph/0301236} \BibitemShut {NoStop}%
\bibitem [{\citenamefont {Wakamatsu}(2009)}]{Wakamatsu:2009fn}%
  \BibitemOpen
  \bibfield  {author} {\bibinfo {author} {\bibfnamefont {M.}~\bibnamefont {Wakamatsu}},\ }\href {https://doi.org/10.1103/PhysRevD.79.094028} {\bibfield  {journal} {\bibinfo  {journal} {Phys. Rev. D}\ }\textbf {\bibinfo {volume} {79}},\ \bibinfo {pages} {094028} (\bibinfo {year} {2009})},\ \Eprint {https://arxiv.org/abs/0903.1886} {arXiv:0903.1886 [hep-ph]} \BibitemShut {NoStop}%
\bibitem [{\citenamefont {Kim}\ and\ \citenamefont {Weiss}(2024)}]{Kim:2023pll}%
  \BibitemOpen
  \bibfield  {author} {\bibinfo {author} {\bibfnamefont {J.-Y.}\ \bibnamefont {Kim}}\ and\ \bibinfo {author} {\bibfnamefont {C.}~\bibnamefont {Weiss}},\ }\href {https://doi.org/10.1016/j.physletb.2023.138387} {\bibfield  {journal} {\bibinfo  {journal} {Phys. Lett. B}\ }\textbf {\bibinfo {volume} {848}},\ \bibinfo {pages} {138387} (\bibinfo {year} {2024})},\ \Eprint {https://arxiv.org/abs/2310.16890} {arXiv:2310.16890 [hep-ph]} \BibitemShut {NoStop}%
\bibitem [{\citenamefont {Kim}\ and\ \citenamefont {Weiss}(2025)}]{Kim:2024ibz}%
  \BibitemOpen
  \bibfield  {author} {\bibinfo {author} {\bibfnamefont {J.-Y.}\ \bibnamefont {Kim}}\ and\ \bibinfo {author} {\bibfnamefont {C.}~\bibnamefont {Weiss}},\ }\href {https://doi.org/10.1103/PhysRevD.111.074007} {\bibfield  {journal} {\bibinfo  {journal} {Phys. Rev. D}\ }\textbf {\bibinfo {volume} {111}},\ \bibinfo {pages} {074007} (\bibinfo {year} {2025})},\ \Eprint {https://arxiv.org/abs/2411.17634} {arXiv:2411.17634 [hep-ph]} \BibitemShut {NoStop}%
\bibitem [{\citenamefont {Kim}(2025)}]{Kim:2025mol}%
  \BibitemOpen
  \bibfield  {author} {\bibinfo {author} {\bibfnamefont {J.-Y.}\ \bibnamefont {Kim}},\ }\href {https://doi.org/10.1103/qpzb-1skm} {\bibfield  {journal} {\bibinfo  {journal} {Phys. Rev. D}\ }\textbf {\bibinfo {volume} {112}},\ \bibinfo {pages} {074032} (\bibinfo {year} {2025})},\ \Eprint {https://arxiv.org/abs/2506.21013} {arXiv:2506.21013 [hep-ph]} \BibitemShut {NoStop}%
\bibitem [{\citenamefont {Minamiguchi}\ and\ \citenamefont {Uji}(2026)}]{Minamiguchi:2026uaq}%
  \BibitemOpen
  \bibfield  {author} {\bibinfo {author} {\bibfnamefont {J.}~\bibnamefont {Minamiguchi}}\ and\ \bibinfo {author} {\bibfnamefont {T.}~\bibnamefont {Uji}},\ }\href@noop {} {\  (\bibinfo {year} {2026})},\ \Eprint {https://arxiv.org/abs/2609.25867} {arXiv:2609.25867 [hep-ph]} \BibitemShut {NoStop}%
\bibitem [{\citenamefont {Minamiguchi}(2026)}]{Minamiguchi:2026qcb}%
  \BibitemOpen
  \bibfield  {author} {\bibinfo {author} {\bibfnamefont {J.}~\bibnamefont {Minamiguchi}},\ }\href@noop {} {\  (\bibinfo {year} {2026})},\ \Eprint {https://arxiv.org/abs/2609.37354} {arXiv:2609.37354 [hep-ph]} \BibitemShut {NoStop}%
\bibitem [{\citenamefont {Kubota}\ \emph {et~al.}(1999)\citenamefont {Kubota}, \citenamefont {Wakamatsu},\ and\ \citenamefont {Watabe}}]{Kubota:1999hx}%
  \BibitemOpen
  \bibfield  {author} {\bibinfo {author} {\bibfnamefont {T.}~\bibnamefont {Kubota}}, \bibinfo {author} {\bibfnamefont {M.}~\bibnamefont {Wakamatsu}},\ and\ \bibinfo {author} {\bibfnamefont {T.}~\bibnamefont {Watabe}},\ }\href {https://doi.org/10.1103/PhysRevD.60.014016} {\bibfield  {journal} {\bibinfo  {journal} {Phys. Rev. D}\ }\textbf {\bibinfo {volume} {60}},\ \bibinfo {pages} {014016} (\bibinfo {year} {1999})},\ \Eprint {https://arxiv.org/abs/hep-ph/9902329} {arXiv:hep-ph/9902329} \BibitemShut {NoStop}%
\bibitem [{\citenamefont {Belitsky}\ \emph {et~al.}(2003)\citenamefont {Belitsky}, \citenamefont {Ji},\ and\ \citenamefont {Yuan}}]{Belitsky:2002sm}%
  \BibitemOpen
  \bibfield  {author} {\bibinfo {author} {\bibfnamefont {A.~V.}\ \bibnamefont {Belitsky}}, \bibinfo {author} {\bibfnamefont {X.}~\bibnamefont {Ji}},\ and\ \bibinfo {author} {\bibfnamefont {F.}~\bibnamefont {Yuan}},\ }\href {https://doi.org/10.1016/S0550-3213(03)00121-4} {\bibfield  {journal} {\bibinfo  {journal} {Nucl. Phys. B}\ }\textbf {\bibinfo {volume} {656}},\ \bibinfo {pages} {165} (\bibinfo {year} {2003})},\ \Eprint {https://arxiv.org/abs/hep-ph/0208038} {arXiv:hep-ph/0208038} \BibitemShut {NoStop}%
\bibitem [{\citenamefont {Adkins}\ \emph {et~al.}(1983)\citenamefont {Adkins}, \citenamefont {Nappi},\ and\ \citenamefont {Witten}}]{Adkins:1983ya}%
  \BibitemOpen
  \bibfield  {author} {\bibinfo {author} {\bibfnamefont {G.~S.}\ \bibnamefont {Adkins}}, \bibinfo {author} {\bibfnamefont {C.~R.}\ \bibnamefont {Nappi}},\ and\ \bibinfo {author} {\bibfnamefont {E.}~\bibnamefont {Witten}},\ }\href {https://doi.org/10.1016/0550-3213(83)90559-X} {\bibfield  {journal} {\bibinfo  {journal} {Nucl. Phys. B}\ }\textbf {\bibinfo {volume} {228}},\ \bibinfo {pages} {552} (\bibinfo {year} {1983})}\BibitemShut {NoStop}%
\bibitem [{\citenamefont {Wakamatsu}\ and\ \citenamefont {Yoshiki}(1991)}]{Wakamatsu:1990ud}%
  \BibitemOpen
  \bibfield  {author} {\bibinfo {author} {\bibfnamefont {M.}~\bibnamefont {Wakamatsu}}\ and\ \bibinfo {author} {\bibfnamefont {H.}~\bibnamefont {Yoshiki}},\ }\href {https://doi.org/10.1016/0375-9474(91)90263-6} {\bibfield  {journal} {\bibinfo  {journal} {Nucl. Phys. A}\ }\textbf {\bibinfo {volume} {524}},\ \bibinfo {pages} {561} (\bibinfo {year} {1991})}\BibitemShut {NoStop}%
\bibitem [{\citenamefont {Kim}\ \emph {et~al.}(2026)\citenamefont {Kim}, \citenamefont {Won},\ and\ \citenamefont {Weiss}}]{Kim:2026gat}%
  \BibitemOpen
  \bibfield  {author} {\bibinfo {author} {\bibfnamefont {J.-Y.}\ \bibnamefont {Kim}}, \bibinfo {author} {\bibfnamefont {H.-Y.}\ \bibnamefont {Won}},\ and\ \bibinfo {author} {\bibfnamefont {C.}~\bibnamefont {Weiss}},\ }\href {https://doi.org/10.1016/j.physletb.2026.140980} {\bibfield  {journal} {\bibinfo  {journal} {Phys. Lett. B}\ }\textbf {\bibinfo {volume} {882}},\ \bibinfo {pages} {140980} (\bibinfo {year} {2026})},\ \Eprint {https://arxiv.org/abs/2605.23125} {arXiv:2605.23125 [hep-ph]} \BibitemShut {NoStop}%
\bibitem [{\citenamefont {Burkardt}(2013)}]{Burkardt:2012sd}%
  \BibitemOpen
  \bibfield  {author} {\bibinfo {author} {\bibfnamefont {M.}~\bibnamefont {Burkardt}},\ }\href {https://doi.org/10.1103/PhysRevD.88.014014} {\bibfield  {journal} {\bibinfo  {journal} {Phys. Rev. D}\ }\textbf {\bibinfo {volume} {88}},\ \bibinfo {pages} {014014} (\bibinfo {year} {2013})},\ \Eprint {https://arxiv.org/abs/1205.2916} {arXiv:1205.2916 [hep-ph]} \BibitemShut {NoStop}%
\bibitem [{\citenamefont {Fukushima}\ and\ \citenamefont {Uji}(2026)}]{Fukushima:2026wwc}%
  \BibitemOpen
  \bibfield  {author} {\bibinfo {author} {\bibfnamefont {K.}~\bibnamefont {Fukushima}}\ and\ \bibinfo {author} {\bibfnamefont {T.}~\bibnamefont {Uji}},\ }\href {https://doi.org/10.1103/fgxk-81p3} {\bibfield  {journal} {\bibinfo  {journal} {Phys. Rev. D}\ }\textbf {\bibinfo {volume} {113}},\ \bibinfo {pages} {114049} (\bibinfo {year} {2026})},\ \Eprint {https://arxiv.org/abs/2603.11704} {arXiv:2603.11704 [hep-ph]} \BibitemShut {NoStop}%
\bibitem [{\citenamefont {Wakamatsu}(2010)}]{Wakamatsu:2010nsh}%
  \BibitemOpen
  \bibfield  {author} {\bibinfo {author} {\bibfnamefont {M.}~\bibnamefont {Wakamatsu}},\ }\href {https://doi.org/10.1140/epja/i2010-10954-6} {\bibfield  {journal} {\bibinfo  {journal} {Eur. Phys. J. A}\ }\textbf {\bibinfo {volume} {44}},\ \bibinfo {pages} {297} (\bibinfo {year} {2010})},\ \Eprint {https://arxiv.org/abs/0908.0972} {arXiv:0908.0972 [hep-ph]} \BibitemShut {NoStop}%
\bibitem [{\citenamefont {Hatta}\ and\ \citenamefont {Schoenleber}(2024)}]{Hatta:2024otc}%
  \BibitemOpen
  \bibfield  {author} {\bibinfo {author} {\bibfnamefont {Y.}~\bibnamefont {Hatta}}\ and\ \bibinfo {author} {\bibfnamefont {J.}~\bibnamefont {Schoenleber}},\ }\href {https://doi.org/10.1007/JHEP09(2024)154} {\bibfield  {journal} {\bibinfo  {journal} {JHEP}\ }\textbf {\bibinfo {volume} {09}},\ \bibinfo {pages} {154}},\ \Eprint {https://arxiv.org/abs/2404.18872} {arXiv:2404.18872 [hep-ph]} \BibitemShut {NoStop}%
\bibitem [{\citenamefont {Kahana}\ and\ \citenamefont {Ripka}(1984)}]{Kahana:1984be}%
  \BibitemOpen
  \bibfield  {author} {\bibinfo {author} {\bibfnamefont {S.}~\bibnamefont {Kahana}}\ and\ \bibinfo {author} {\bibfnamefont {G.}~\bibnamefont {Ripka}},\ }\href {https://doi.org/10.1016/0375-9474(84)90692-4} {\bibfield  {journal} {\bibinfo  {journal} {Nucl. Phys. A}\ }\textbf {\bibinfo {volume} {429}},\ \bibinfo {pages} {462} (\bibinfo {year} {1984})}\BibitemShut {NoStop}%
\bibitem [{\citenamefont {Varshalovich}\ \emph {et~al.}(1988)\citenamefont {Varshalovich}, \citenamefont {Moskalev},\ and\ \citenamefont {Khersonskii}}]{Varshalovich:1988ifq}%
  \BibitemOpen
  \bibfield  {author} {\bibinfo {author} {\bibfnamefont {D.~A.}\ \bibnamefont {Varshalovich}}, \bibinfo {author} {\bibfnamefont {A.~N.}\ \bibnamefont {Moskalev}},\ and\ \bibinfo {author} {\bibfnamefont {V.~K.}\ \bibnamefont {Khersonskii}},\ }\href {https://doi.org/10.1142/0270} {\emph {\bibinfo {title} {{Quantum Theory of Angular Momentum}: {Irreducible Tensors, Spherical Harmonics, Vector Coupling Coefficients, 3nj Symbols}}}}\ (\bibinfo  {publisher} {World Scientific Publishing Company},\ \bibinfo {year} {1988})\BibitemShut {NoStop}%
\end{thebibliography}%

\end{document}